\documentstyle[12pt]{article}

\newcommand{\nn}{\nonumber \\}

\newcommand{\R}{R}
\newcommand{\C}{C}
\newcommand{\Z}{Z}
\newcommand{\arr}{\left(\begin{array}{cc}}
\newcommand{\ay}{\end{array}\right)}
\newcommand{\inv}{^{-1}}
\newcommand{\parc}{\partial}
\newcommand{\k}{\underline{k}}
\newcommand{\x}{\underline{x}}

\def\lbq#1{\label{eq:#1}}
\def\rrf#1{(\ref{eq:#1})}
\begin{document}

\title
{\begin{flushright}
\normalsize{ITP Budapest Report 509}
\normalsize{\\ hep-th 9502145}
\end{flushright}
\vspace{2cm}
\bf Reduced ${\bf SL(2,\R)}$ WZNW Quantum Mechanics\thanks{To appear in
{\it J. Math. Phys.}}} \author{ T. F\"ul\"op\thanks{E-mail:
fulopt@hal9000.elte.hu}
      \\ {\it Institute for Theoretical Physics} \\
         {\it E\"otv\"os University, Budapest, Hungary}}
\date{20 November 1995}
\maketitle

    \begin{abstract}

The $SL(2,\R)$ WZNW $\rightarrow$ Liouville reduction leads to a nontrivial
phase space on the classical level both in $0+1$ and $1+1$ dimensions. To
study the consequences in the quantum theory, the quantum mechanics of the
$0+1$ dimensional, point particle version of the constrained WZNW model is
investigated. The spectrum and the eigenfunctions of the obtained---rather
nontrivial---theory are given, and the physical connection between the pieces
of the reduced configuration space is discussed in all the possible cases of
the constraint parameters.
    \end{abstract}


    \section{Introduction}\label{intro}

In the past several years the Toda models have been studied intensively. In
these field theories scalar fields are coupled to each other by certain
special exponential terms, in a way that corresponds to a simple Lie algebra.
The Toda models can be considered as generalizations of the Liouville theory,
which is of particular interest since it appears in many problems of physics
and mathematics. An interesting means of deriving and studying the remarkable
properties---integrability, conformal invariance and W-algebraic symmetry---of
the Toda models is offered by the observation that the Toda theories can be
obtained by a suitable reduction of the Wess-Zumino-Novikov-Witten (WZNW)
model \cite{FL2}. The WZNW model is a theory of a field that takes its values
from a Lie group G, and the reduction procedure - by imposing appropriate
(first class) constraints - associates it to a Toda theory that corresponds to
the Lie algebra of G. In the case $G=SL(2,R)$ the reduced theory is nothing
but the Liouville theory.

However, the connection between the WZNW and Toda models is more intricate. A
closer look at the reduction procedure shows that it yields not exactly the
Toda theory but a richer structure, the Toda theory arises only as a
component, a subsystem of it. This aspect was first noticed in \cite{FL4}. To
study the precise relation of the Toda models to the WZNW ones a recent work
examined the $SL(2,R)$ WZNW $\rightarrow$ Liouville reduction from the phase
space point of view \cite{1}. The authors considered the classical $SL(2,\R)$
WZNW model, imposed the appropriate constraints and described the reduction of
the phase space under the constraints. They found that the reduced phase space
contains two subsystems (nonintersecting open submanifolds) that admit a clear
physical interpretation. On both subsets the reduced WZNW theory leads to the
Liouville theory locally, but these two copies of Liouville theories are not
independent. The connection between them comes from a 'border line', a lower
dimensional surface in the reduced phase space connecting them. For a better
understanding of the situation \cite{1} carried out a similar analysis on the
$0 + 1$ dimensional, point mechanical analogue of the $SL(2,\R)$ WZNW model.
This can be thought as the space independent version, the 'zero mode sector'
of the WZNW model. In \cite{1} it was found that depending on the signs of the
constraint parameters one can arrive at two different types of reduced
theories. In both cases the phase space reduces into two locally independent
parts. The difference is that when the constraint parameters have equal signs
the two halves are disconnected, there is no 'border line' between them. As a
result a classical motion cannot touch both parts. Actually, though the
reduced Hamiltonian is not of the usual form of the sum of a kinetic and a
potential term, the system behaves as if the two halves of the reduced one
dimensional configuration space were separated by an infinitely high potential
barrier. On the other hand, if the signs of the constraint parameters are
opposite then an analogue of an infinitely deep potential well characterises
the situation. In this case the two 'half worlds' are connected, a motion can
cross the point that separates the two halves of the reduced configuration
space, moreover, the negative energy motions will oscillate between the two
parts. \cite{1} concludes that it is not enough to give the (global) reduced
theory in the local coordinates of the reduced phase space. The arising two
components of the reduced theory seem to be independent while actually they
may have a physical connection, a correlated behaviour which can be discovered
only from the global point of view. (\cite{1} also contains results about the
$SL(n,\R)$ point particle model and discovers some similar properties for
$n>2$ as well.)

These are the features of the classical theories. What happens on the quantum
level? Because of the expected difficulties of quantizing the $SL(2,\R)$ WZNW
model it is useful to examine the point particle version. Turning to the
masspoint theory the following questions arise naturally: what kind of
relation connects the two half-worlds in the quantum theory? Are there any
oscillating motions quantum mechanically when the the constraint parameters
are of opposite sign? Similarly we may ask if in the case of coinciding signs
the separation remains or a tunneling is allowed. Moreover, in the first case
one may expect negative energy bound states. Are they really present? In this
paper we solve the quantum theory of the masspoint version, and thus we can
answer whether our expectations based on the classical behaviour hold. While
---as a consequence of its rather nontrivial characteristics---the point
particle problem is interesting in itself, its properties are expected to shed
a light on the field theoretical case, just as it happened on the classical
level \cite{1}.

The paper is organized as follows. Sect.\ 2 and 3 present the classical
mechanics of the $SL(2,\R)$ masspoint before resp. after imposing the
point mechanical form of the WZNW $\rightarrow$ Liouville constraints. The
quantum mechanics of the unconstrained and the constrained systems are
established in Sect.\ 4 and 5. The reduced system splits into two parts in
a symmetric way, Sect.\ 6 gives the eigenfunctions on one such part.
Sect.\ 7 investigates the orthogonality and the completeness properties of
these 'half-eigenfunctions'.  The connection between the two parts is
examined in Sect.\ 8. Sect.\ 9 discusses the coordinate independence of
the definition of the constrained system.  Sect.\ 10 gives the conclusions
of the paper. The larger proofs and calculations belonging to the
statements of Sect.\ 7 are presented in two appendices, one concerning
orthogonality and one concerning completeness.

    \section{The classical mechanics of the unconstrained
theory}\label{3dcm}

The $SL(2,\R)$ WZNW model is defined by the following action:
    \begin{eqnarray}
    S_0 = \frac{m}{4\pi} \int d\sigma d\tau
    \, \mbox{Tr} \! \left[ (g^{-1} \partial_\tau g)^2 -
    (g^{-1} \partial_\sigma g)^2 \right] + \nn
    m' \int d^3 x \: \epsilon^{ijk} \, \mbox{Tr} \! \left( g^{-1}
    \partial_i g \: g^{-1} \partial_j g \: g^{-1} \partial_k g \right),
    \lbq{10}
    \end{eqnarray}
where $\sigma$ and $\tau$ coordinate a two dimensional Minkowski space, $g$ is
an $SL(2,\R)$-valued function of $\sigma$ and $\tau$ being periodic in
$\sigma$ with period $2\pi$. The coefficients of the first term, the action of
the $SL(2,\R)$ sigma model, and the second, topological term called
Wess-Zumino term, are denoted by $m/{4\pi}$ and $m'$ respectively. The point
particle version, i.e. the physics of the zero modes of the field theory
\rrf{10} is defined by restricting the configurations to the space-independent
ones $g = g(\tau)$ only. Then the action reduces to
    \begin{equation}
    S = \frac{m}{2} \int d\tau \, \mbox{Tr} \!
    \left[ (g^{-1} \partial_\tau g)^2 \right] .
    \lbq{20}\end{equation}
The motion of the point particle is a function $g: \R \rightarrow SL(2,\R)$;
we can see that the Wess-Zumino term does not contribute to the masspoint
version. The theory posesses left and right translation symmetries under the
transformations $ g \mapsto hg $, $ g \mapsto gh^{-1} $, $ h \in SL(2,\R) $,
the corresponding conserved quantities are $ J = \dot g g^{-1} $ and $ \tilde
J := g^{-1} \dot g $, taking their values in $sl(2,\R)$, the Lie algebra of
$SL(2,\R)$.

The equation of motion following from the action \rrf{20} is
$ (g^{-1} \dot g)
\dot{\,} = 0 $ (or, equivalently, $ (\dot g g^{-1}) \dot{\,} = 0 $).
Its solution is $ g(t) = g(0) \, e^{At} $, where $ A \in sl(2,\R)
$ is a kind of initial data specified by the initial conditions as $ A =
g(0)^{-1} \dot g(0) $.
Using the basis $T_1 = \sigma_1$, $T_2 = i \sigma_2$, $T_3 = \sigma_3$ in
$sl(2,\R)$ (where $\sigma_k$-s denote the Pauli-matrices) $A$ can be written
as $A^k T_k$ (Einstein-summation). We will make use of the identity
    \begin{equation} e^{c^k
    \sigma_k} = \cosh R \: {\bf 1} + \sinh R /R \: c^k \sigma_k
    \lbq{60}\end{equation}
where $R^2 = \frac{1}{2} \mbox{Tr} [(c^k \sigma_k)^2]$ (this formula also
holds for $\mbox{Tr} [(c^k \sigma_k)^2] < 0$ with $R$ being imaginary and for
$\mbox{Tr} [(c^k \sigma_k)^2] = 0$ with $\cosh R \rightarrow 1, \sinh R /R
\rightarrow 1$). Based on \rrf{60} the solution of the equation of motion is
    \begin{equation}
    g(t) = g(0) \left( \begin{array}{cc}
    \cosh (rt) + A^3 \sinh (rt) /r & (A^1 + A^2) \sinh (rt) /r \\
    (A^1 - A^2) \sinh (rt) /r & \cosh (rt) - A^3 \sinh (rt) /r
    \end{array} \right)
    \lbq{70}\end{equation}
with $r^2 = \frac{1}{2} \mbox{Tr} [(A^k T_k)^2] $ (also allowing zero or
imaginary values of $r$). If the trace is negative then the solution also
can be expressed as
    \begin{equation}
    g(t) = g(0) \left( \begin{array}{cc}
    \cos (\omega t) + A^3 \sin (\omega t) / \omega & (A^1 + A^2)
    \sin (\omega t) /
    \omega \\ (A^1 - A^2) \sin (\omega t) / \omega & \cos (\omega t) -
    A^3 \sin
    (\omega t) / \omega \end{array} \right)
    \lbq{73}\end{equation}
with $\omega = \sqrt{-r^2}$. For $r = 0$ the solution actually reads
    \begin{equation}
    g(t) = g(0) \left( \begin{array}{cc}
                1 + A^3t     & (A^1 + A^2)t \\
                (A^1 - A^2)t & 1 - A^3t
    \end{array} \right).
    \lbq{76}\end{equation}
A compact form of formula \rrf{70} is $ g(t) = g(0) \, ( \cosh(rt) \, {\bf 1}
+ \sinh(rt) / r \, A ) $. We can see that the $r^2 < 0$ solutions are closed
because of their trigonometrical time dependence, the $r^2 \geq 0$ motions are
open.

To study the canonical structure of the theory let us consider a
parametrization of $SL(2,\R)$ $ g(\xi^i), \: i = 1, 2, 3 $. The Lagrangian
reads as
    \begin{equation}
    L(\xi, \dot{\xi}) = \frac{m}{2} h_{kl} (\xi) \: \dot{\xi^k} \dot{\xi^l},
    \lbq{80}\end{equation}
where $ h_{kl}(\xi) = \mbox{Tr} \left[ g^{-1} \partial_k g g^{-1} \partial_l
g \right] $ is the metric tensor on $SL(2,\R)$. The canonical momenta are
    \begin{equation}
    p_k = \frac{\partial L}{\partial \dot{\xi^k}} = m h_{kl} \dot{\xi^l},
    \lbq{100}\end{equation}
from which we get the Hamiltonian
    \begin{equation}
    {\cal H} (\xi, p) = \frac{1}{2m} h^{kl} p_k p_l
    \lbq{110}\end{equation}
($\{h^{kl}\}$, the inverse of the matrix $\{h_{kl}\}$ exists because
$SL(2,\R)$ is simple).

The Hamiltonian can be expressed as the function of $\xi$ and $\dot{\xi}$
as well. One finds that
    \begin{equation}
    {\cal H}
    (\xi, \dot{\xi}) = \frac{m}{2} h_{kl}(\xi) \: \dot{\xi^k} \dot{\xi^l} =
    L(\xi, \dot{\xi}).
    \lbq{115}\end{equation}
Combining this result with the coordinate free form of the Lagrangian (cf. eq.
\rrf{20}) and $ g(t) = g(0) \, e^{At} $, the value of the
Hamiltonian on a solution of
the equation of motion can be determined easily:
    \begin{equation}
    {\cal H} = \frac{m}{2} \mbox{Tr} [A^2] = m r^2.
    \lbq{117}\end{equation}
As a consequence we can see that the energy is negative for a closed motion
and nonnegative for open motions.

Let us turn to a special parametrization, namely the one which is based on the
Gauss decomposition:
    \begin{equation}
    \left( \begin{array}{cc} \alpha & \beta \\ \gamma & \delta \end{array}
    \right) = \left( \begin{array}{cc} 1 & a \\ 0 & 1 \end{array} \right)
    \left( \begin{array}{cc} 1/\delta & 0 \\ 0 & \delta \end{array} \right)
    \left( \begin{array}{cc} 1 & 0 \\ c & 1 \end{array} \right).
    \lbq{120}\end{equation}
This parametrization describes any $\left( \begin{array}{cc} \alpha & \beta \\
\gamma & \delta \end{array} \right) \in SL(2,\R)$ except those having $\delta
= 0$. With $\delta = \xi^1$, $a = \xi^2$ and $c = \xi^3$ $\, \{h_{kl}\}$ is of
the form
    \begin{equation}
    \{h_{kl}\} = \left( \begin{array}{ccc} 2/\delta^2 & 0 & 0 \\
    0 & 0 & \delta^2 \\ 0 & \delta^2 & 0 \end{array} \right).
    \lbq{130}\end{equation}
The determinant of this metric is
    \begin{equation}
    h = -2 \delta^2,
    \lbq{140}\end{equation}
and the inverse of the metric tensor, $\{h^{kl}\}$ has the form
    \begin{equation}
    \{h^{kl}\} = \left( \begin{array}{ccc} \delta^2/2 & 0 & 0 \\
    0 & 0 & 1/\delta^2 \\ 0 & 1/\delta^2 & 0 \end{array} \right).
    \lbq{150}\end{equation}
Thus we arrive at the following Hamiltonian in this parametrization:
    \begin{equation}
    {\cal H} (\delta, a, c, p_\delta, p_a, p_c) = \frac{1}{4m} \delta^2
    p_\delta^2 + \frac{1}{m} \frac{p_a p_c}{\delta^2}.
    \lbq{160}\end{equation}
Expression \rrf{160} shows that $a$ and $c$ are cyclic coordinates since $\cal
H$ is independent of them. This is an advantage of using the parameters
$\delta, a, c$. Later we will see that this parametrization fits very well for
our further considerations. That's why in the following we will work in these
coordinates.

    \section{The constrained model on the classical level}\label{1dcm}

Now we impose the constraints that reduce the $SL(2,\R)$ WZNW theory to the
Liouville one. For space independent configurations they read
    \begin{equation}
    \mbox{Tr} \left[ e_{12} \, \dot g g\inv \right] = \mu, \qquad
    \mbox{Tr} \left[ e_{21} \, g\inv \dot g \right] = \nu,
    \lbq{170}\end{equation}
where $ e_{12} = \arr 0 & 1 \\ 0 & 0 \ay $ and $ e_{21} = \arr 0 & 0 \\ 1 & 0
\ay $. Expressing \rrf{170} in the $\delta, a, c$ parametrization gives
    \begin{equation}
    p_a = m \mu, \qquad p_c = m \nu.
    \lbq{190}\end{equation}
As $p_a$ and $p_c$ are constants of the unrestricted motion we see that the
constraints mean nothing else but a special choice of some of the initial
conditions. This is a general feature of first class constraints, that these
constants of motion happen to be canonical momenta is the real advantage of
the parameters $\delta, a, c$.

The reduced phase space can be obtained by factorizing the complete phase
space by the gauge transformations these first class constraints generate. The
action of these gauge transformations is
    \begin{equation}
    g \mapsto e^{\theta_L e_{12}} g \: e^{\theta_R e_{21}} = \arr 1 &
    \theta_L \\ 0 & 1 \ay g \arr 1 & 0 \\
    \theta_R & 1 \ay
    \lbq{210}\end{equation}
(see \cite{1}) where $\theta_L$ and $\theta_R$ are the two parameters of the
transformations. This symmetry transformation acts on $a$ and $c$ as $a
\mapsto a + \theta_L$, $c \mapsto c + \theta_R$, while it leaves
$\delta$, $p_\delta$, $p_a$ and $p_c$ invariant. Thus the factorization simply
means that $\delta$ and $p_\delta$ parametrize the reduced phase space.

Any motion $\delta(t), p_\delta(t)$ allowed by the constrained dynamics
corresponding to the constraint parameters $\mu$ and $\nu$ can be obtained by
taking a solution $\delta(t), a(t), c(t)$, $p_\delta(t), p_a, p_c$ of the
unconstrained theory where $p_a = m \mu$, $p_c = m \nu$. Roughly speaking we
just have to omit $a(t)$ and $c(t)$. The coordinate $\delta \in \R$ survives
the reduction, thus one may consider it as the coordinate of the one
dimensional configuration space of the reduced theory. The constrained
dynamics is governed by the Hamiltonian
    \begin{equation}
    H(\delta, p_\delta) = \frac{1}{4m} \delta^2 p_\delta^2 +
    ms \frac{1}{\delta^2},
    \lbq{230}\end{equation}
where $s = \mu \nu$.

Until now we have found the Gauss decomposition a very appropriate way of
introducing coordinates on the group $SL(2,\R)$ to reach the canonical
structure of the reduced system. Now let us face the problematic side of this
approach.

The topology of $SL(2,\R)$ is not $\R^3$ but $ \R^2 \times S^1 $, consequently
it cannot be covered by a single parametrization. In the case of the Gauss
decomposition the signal of this is that the Gauss decomposition works
only for
$\delta \neq 0$, the elements $\arr \alpha & \beta \\ \gamma & 0 \ay$ are left
out. As a result the canonical coordinates $\delta$, $a$, $c$, $p_\delta$,
$p_a$ and $p_c$ parametrize only two nonintersecting open submanifolds, two
'open halves' of the whole phase space, corresponding to the two regions
$-\infty < \delta < 0$ and $0 < \delta < \infty$. In particular, the point $
\delta = 0 $ is left out from the reduced configuration space. In the
meantime, extracted from \rrf{70}, the time dependence of $\delta$ is of the
form $ \delta(t) = C_1 \sinh(rt)/r + C_2 \cosh(rt) $ where $C_1$ and $C_2$ are
arbitrary constants. Thus we can see that there exist motions that cross
the $
\delta = 0 $ surface in the whole phase space---for example for imaginary
values of r oscillations occur between the regions $ \delta > 0 $ and $ \delta
< 0 $. Furthermore, for such a motion $p_\delta$ tends to infinity as the
particle reaches $ \delta = 0 $. This can be seen from $ p_\delta(t) = 2m \dot
\delta / \delta^2 $, the connection between $p_\delta$ and the smoothly
varying quantity $ \dot \delta $ (cf. \rrf{100} and \rrf{130}). Hence the
coordinates $\delta$, $p_\delta$ of the reduced phase space seem to be able to
describe only those parts of a motion when $ \delta(t) \neq 0 $, they cannot
give an account of how the particle moves across $ \delta = 0 $.

Fortunately we can overcome these difficulties. First, let us complete the
reduced configuration space by mapping the points $\arr \alpha & \beta \\
\gamma & 0 \ay$ of the configuration space of the unconstrained system to the
point $\delta = 0$ of the reduced configuration space, just as we have mapped
the points $\arr \alpha & \beta \\ \gamma & \delta \neq 0 \ay$ to $\delta$.
Second, let the recipe to follow a motion through $ \delta = 0 $ be to fit the
quantity $ \delta^2 p_\delta $. This recipe is clear from the unconstrained
point of view: we simply fit $\dot \delta$ here. By these tricks we can
eliminate the disadvantage of working with only one parametrization instead of
covering $SL(2,\R)$ with more than one patches. Clearly, the canonical
formalism of the unconstrained system also owes the problem of the missing $
\delta = 0 $. Nevertheless, the solution given here for the reduced system
applies in a straightforward way for the unconstrained one, too.

We will see that the $\delta = 0$ problem also arises in the quantum theory.
There it appears as an irregular singularity of the Hamiltonian at $\delta =
0$ and the challenge will be to define the quantum theory on the whole
configuration space despite this singularity.

Finally let us introduce a canonical transformation which transforms the
constrained Hamiltonian to a form of a sum of a kinetic and a potential term.
This can be achieved by the following transformation:
    \begin{equation}
    x := \sqrt{2} \ln \delta,
    \lbq{280}\end{equation}
    \begin{equation}
    p_x := \frac{1}{\sqrt{2}} \delta p_\delta.
    \lbq{290}\end{equation}
The resulting Hamiltonian is
    \begin{equation}
    H_x (x, p_x) = \frac{1}{2m} p_x^2 + m s e^{-\sqrt{2}x}.
    \lbq{300}\end{equation}
The price we have to pay for having such a nice Hamiltonian is that by
\rrf{280} we restricted ourselves to $\delta > 0$ only. (Or, because of the
$\delta \leftrightarrow -\delta$ symmetry of the system, to $\delta < 0$, if
writing $-\delta$ instead of $\delta$ in \rrf{280}.) Remarkably, the
logarithmic connection between $x$ and $\delta$ is the point mechanical
analogue of the one that
relates the field of the reduced SL(2,R) WZNW theory to the Liouville field
$\phi$ in the field theoretical case \cite{FL4}.

With the aid of \rrf{300} it is easy to analyse the three qualitatively
different situations arising. If $s > 0$ then the potential increases
exponentially as we travel to the negative $x$ direction. Thus for all the
allowed motions with positive energies there is a turning point when moving to
the negative direction towards $x = -\infty$ $(\delta = 0)$. In this case
there is no possibility for the masspoint to cross the border $\delta = 0$.
For $s = 0$ we have a free particle moving along the $x$ axis. Now the 'point'
$x = -\infty$ $(\delta = 0)$ cannot be reached in a finite time interval so
the masspoint cannot cross the border even in this case. In the case $s < 0$
an exponentially deep potential valley attracts the particle towards the
negative direction, what's more, the time needed to reach $x = -\infty$
happens to be finite. This shows that for $s < 0$ the particle may cross the
border $\delta = 0$.

    \section{The quantum theory of the unconstrained system}\label{3dqm}

Let us define the quantum mechanics of the point particle $SL(2,\R)$ WZNW
theory via canonical quantization. We use the coordinates $\delta, a, c$ and
work in the coordinate representation. The wave functions are then complex
valued functions defined for all $\delta \in \R \setminus \{0\}, a \in \R$ and
$c \in \R$. We define the scalar product as
    \begin{equation}
    (\Psi_1, \Psi_2) := \int \Psi_1^* \Psi_2^{\,} \sqrt{-h} \,
    d\delta \, da \, dc .
    \lbq{420}\end{equation}
Here $h$ denotes the determinant of the matrix $\{h_{kl}\}$ in the $\delta, a,
c$ parametrization (cf. \rrf{140}). The measure in this integral is the usual
one used on curved manifolds. Moreover, in our case the metric tensor is
invariant under the left and right transformations. Consequently, if we
adopt the left and right transformations for the wave functions:
    \begin{equation}
    [ D_L (h) \Psi ] (g) := \Psi (h\inv g), \qquad [ D_R (h) \Psi ] (g) :=
    \Psi (gh), \qquad g, h \in SL(2,\R),
    \lbq{422}\end{equation}
then the scalar product is also invariant. This property is inevitable if
we want the left and right symmetries of the classical theory to be present on
the quantum level as well. Observe that these natural requirements led to the
appearance of a nontrivial weight function $\rho(\delta) := \sqrt{-h} =
\sqrt{2} \vert\delta\vert$ in the integral \rrf{420}.

In coordinate representation the canonical momenta are defined as partial
derivations with respect to the corresponding coordinates if the
configuration space is flat. For curved configuration spaces this definition
does not give symmetric operators because of the presence of the weight
function in the scalar product. Symmetricity and the requirement $ [ \hat
\xi_k, \hat p_l ] = i \hbar \, \delta_{kl} $ meet in the definition $ \hat
p_k
:= \frac{\hbar}{i} ( \parc_k + \frac{1}{2} \parc_k \ln \sqrt{-h} ) $ (see
\cite{DESY} for example). In our case this gives $ \hat{p}_\delta :=
\frac{\hbar}{i} ( \partial_\delta + \frac{1}{2 \delta} ) $, $ \hat{p}_a :=
\frac{\hbar}{i} \partial_a $ and $ \hat{p}_c := \frac{\hbar}{i} \partial_c $.

Now the stage is set for the definition of the Hamiltonian. This is also a step
which needs some care. The problem with the naive quantum analogue of the
classical Hamiltonian \rrf{160}:
    \begin{equation}
    \hat{\cal H}_n = \frac{1}{4m} \hat \delta^2 \hat p_\delta^2 +
    \frac{1}{m} \hat{p}_a \hat{p}_c \frac{1}{\hat \delta^2}
    \lbq{460}\end{equation}
is that $\hat{\cal H}_n$
is not symmetric (do not forget the weight function!). We
need to find an appropriate ordering of $\hat \delta$ and $\hat p_\delta$.
Moreover, symmetricity is not the only requirement since we would like the
Hamiltonian to be invariant under the left-right symmetry transformations as
well. That is why we turn to the Laplacian of $SL(2,\R)$:
    \begin{equation}
    \triangle \Psi = \frac{1}{\sqrt{-h}} \frac{\partial}{\partial \xi^i} \!
    \left( \sqrt{-h} h^{ij}(\xi) \frac{\partial}{\partial \xi^j} \Psi \right),
    \lbq{470}\end{equation}
the usual definition for curved manifolds, which reads in the case of the
$\delta, a, c$ parametrization
    \begin{equation}
    \triangle \Psi = \frac{1}{2\vert\delta\vert} \, \partial_\delta
    (\vert\delta\vert^3 \, \partial_\delta \Psi) +
    \frac{2}{\delta^2} \partial_a \partial_c \Psi.
    \lbq{480}\end{equation}
The definition \rrf{470} and the properties of the metric tensor guarantee
that the Laplacian is invariant not only under reparametrizations but under
left and right transformations as well. The reason why the Laplacian is
interesting is that $-\frac{\hbar^2}{2m} \triangle$ is just a re-ordered
version of $\hat{\cal H}_n$. So we define the Hamiltonian as
    \begin{equation}
    \left( \hat {\cal H} \Psi \right) (\delta, a, c) := -\frac{\hbar^2}{4m}
    \frac{1}{\vert\delta\vert} \, \partial_\delta \! \left( \vert\delta\vert^3
    \partial_\delta \Psi(\delta, a, c) \right) - \frac{\hbar^2}{m}
    \frac{1}{\delta^2} \partial_a \partial_c \Psi(\delta, a, c) .
    \lbq{490}\end{equation}
A simple calculation shows that this $\hat{\cal H}$ is symmetric (with respect
to the weight function $\rho$). This choice of $\hat{\cal H}$ is the usual and
natural one for curved configuration spaces (see, for example, \cite{DESY}).

We mention that we have not defined the wave functions at $\delta = 0$ and the
operators $\hat \delta$, $\hat{p}_\delta$ and $\hat {\cal H}$ are apparently
ill-defined at $\delta = 0$. These singularities are only coordinate artifacts
here. This won't be the case for the reduced system as we will see soon.

    \section{The quantum mechanics of the reduced system}\label{1dqm}

Being ready with the quantum mechanics of the unconstrained theory the next
task is to consider the quantum analogue of the constraints and see what the
reduction yields. Let us impose the constraints on the quantum level as
    \begin{equation}
    \hat{p}_a \Psi = m \mu \Psi, \qquad \hat{p}_c \Psi = m \nu \Psi
    \lbq{500}\end{equation}
(cf. \rrf{190}). It is very easy to find the wave functions that satisfy
\rrf{500}, they are of the form
    \begin{equation}
    \Psi(\delta, a, c) = \psi(\delta) \, e^{\frac{i}{\hbar} (m \mu a + m
    \nu c)}.
    \lbq{510}\end{equation}
Remarkably, the action of the operators $\hat \delta$, $\hat p_\delta$ and
$\hat {\cal H}$ on a wave function of this form touches only its
$\delta$-depending part. This makes it possible to work with $\psi(\delta)$
instead of $\Psi(\delta, a, c)$ and to consider the one dimensional quantum
mechanics driven by a Hamiltonian $\hat H$:
    \begin{equation}
    (\hat H \psi)(\delta) = -\frac{\hbar^2}{4m}
    \frac{1}{\vert\delta\vert} \, \partial_\delta \! \left(
    \vert\delta\vert^3 \partial_\delta \psi(\delta) \right) +
    m s \frac{1}{\delta^2} \psi(\delta)
    \lbq{540}\end{equation}
together with the scalar product
    \begin{equation}
    (\psi_1,\psi_2) := \int \psi_1^* \psi_2^{\,} \, \rho(\delta)
    \, d\delta
    \lbq{545}\end{equation}
Therefore in the following we investigate the properties of this one
dimensional quantum system. (The classical $H$ corresponding to this
$\hat H$ is just \rrf{230} as we expect.) Here we witness how the
decoupling of the variables $a$, $c$ from the system happens on the quantum
level.

The wave functions \rrf{510} are not square integrable in the $SL(2,\R)$
sense. This is a natural consequence of the constraints that decrease the
degrees of freedom by two, we will require square integrability
'in the reduced sense', i.e. with
respect to the scalar product \rrf{545}. The situation is similar to the
case of a free masspoint in a three dimensional Euclidean space with the
constraints classically imposed as $p_y = \mbox{const.}, p_z = \mbox{const.}$
When we consider the corresponding quantum theory it is obvious that the
normalizability of the wave functions must be understood 'in the one
dimensional sense'.

The one dimensional problem we arrived at is quite an unusual one. The
Hamiltonian is not a
    \begin{equation}
    \psi \mapsto -b^2 \psi'' + V \psi
    \lbq{550}\end{equation}
-type (with a real constant $b$ and a potential function $V$) and a nontrivial
weight function is present in the scalar product. It would be very convenient
if our system could be transformed to an 'ordinary' one with a Hamiltonian of
the form \rrf{550} and with no weight function. For this purpose let us
consider a transformation of $\delta$ to a new variable $x$:
    \begin{equation}
    x = g(\delta)
    \lbq{560}\end{equation}
accompanied by a change of $\psi(\delta)$ to a new wave function $\chi(x)$
which is related to $\psi(\delta)$ as
    \begin{equation}
    \psi(\delta) = f(\delta) \: \chi(g(\delta)),
    \lbq{570}\end{equation}
where $f$ and $g$ are arbitrary real functions. To see how $\hat{H}$
transforms under such a transformation one has to examine the transformation
of the {\em (time dependent) Schr\"odinger equation \/}:
    \begin{equation}
    i \hbar \frac{\parc \psi}{\parc t} (\delta, t) =
    (\hat{H} \psi)(\delta, t).
    \lbq{580}\end{equation}
After a simple calculation the result can be written in the form
    \begin{eqnarray}
    i \hbar \frac{\parc \chi}{\parc t} (x, t) & = & - \frac{\hbar^2}{4m}
    \delta^2 (g')^2 \chi'' - \frac{\hbar^2}{4m} \left( 3 \delta g' + 2 \delta^2
    \frac{f'}{f} g' + \delta^2 g'' \right) \chi' + \nonumber \\ & &
    \left[ -\frac{\hbar^2}{4m} \left( 3 \delta \frac{f'}{f} + \delta^2
    \frac{f''}{f} \right) + m
    s \frac{1}{\delta^2} \right] \chi.
    \lbq{590}\end{eqnarray}
Requiring eq. \rrf{590} to have the form given by eq. \rrf{550} gives a number
of equations. First of all the coefficient of the second derivative must be a
negative constant, for convenience we want to set it to $-\frac{\hbar^2}{2m}$.
This yields
    \begin{equation}
    \delta^2 (g')^2 = 2.
    \lbq{600}\end{equation}
The solution of this condition is
    \begin{equation}
    g(\delta) = \sqrt{2} \ln(c_1 \delta)
    \lbq{610}\end{equation}
where $c_1$
is an arbitrary positive constant; for the moment let us restrict
ourselves to the positive $\delta$ half line. Secondly we require that the
$\chi'$ term must vanish. This and \rrf{610} together yield
    \begin{equation}
    f(\delta) = \frac{c_2}{\delta}
    \lbq{620}\end{equation}
with a nonzero constant $c_2$.
Under the transformation \rrf{570} with such an
$f(\delta)$ and $g(\delta)$ the weight function changes as
    \begin{equation}
    \int_{\delta_1}^{\delta_2} \psi_1^* (\delta) \psi_2^{\,} (\delta)
    \sqrt{2} \vert\delta\vert d\delta = \int_{g(\delta_1)}^{g(\delta_2)}
    \chi_1^* (x) \chi_2^{\,} (x) \, c_2^2 \, dx.
    \lbq{630}\end{equation}
We want to transform away the weight function entirely. This can be reached
simply by choosing $c_2 = 1$.
The parameter $c_1$ does not have such an interesting
specific value, we set $c_1 = 1$. The transformation of the variable
    \begin{equation}
    x = \sqrt{2} \ln \delta
    \lbq{633}\end{equation}
and the wave function
    \begin{equation}
    \psi(\delta) = \frac{1}{\delta} \, \chi(x) = \frac{1}{\delta} \, \chi
    (\sqrt{2} \ln \delta)
    \lbq{636}\end{equation}
lead to the following $\hat{H}_x$:
    \begin{equation}
    (\hat{H}_x \chi)(x) = -\frac{\hbar^2}{2m} \chi''(x) + \left(
    \frac{\hbar^2}{4m} +
    m s e^{-\sqrt{2}x} \right) \chi(x).
    \lbq{640}\end{equation}

This transformation is just the quantum analogue of \rrf{280} and \rrf{300}.
(The additional constant in the potential term of $\hat{H}_x$ is due to the
ordering procedure we maintained at the definition of the quantum
Hamiltonian.) Unfortunately the problem is the same as well: it works only for
the positive half of the configuration space (or the negative one, if
exploiting the symmetry $ \delta \leftrightarrow -\delta $). Nevertheless,
$\hat{H}_x$ will be very useful to understand the physics encoded in
$\hat H$.

    \section{Eigenfunctions on the half configuration
space}\label{felsfv}

With the aid of the transformed Hamiltonian \rrf{640} one can have a rough
picture of the reduced theory. In the cases $s > 0$, $s < 0$ it is more or less
similar to a system with a potential infinitely increasing or decreasing for
$\delta \rightarrow 0$, while for $s = 0$ the system is somehow a 'sum' of two
free theories. The potential valley of the case $s < 0$ suggests to use the
Bohr-Sommerfeld quantization to get a first impression about the spectrum of
$\hat{H}$. For this purpose let's consider a classical motion with
energy $E$ and express $p_\delta$ as a function of $\delta$ using \rrf{230}:
    \begin{equation}
    p_\delta = \pm\frac{\sqrt{4mE \delta^2 - 4 m^2 s}}{\delta^2}.
    \lbq{650}\end{equation}
For the Bohr-Sommerfeld quantization condition the phase space area $\oint
p_\delta d\delta$ is needed as a function of $E$. The problem is that for
$\delta \approx 0$ the behaviour of $p_\delta$ is $p_\delta \sim 1/\delta^2$
(here and in the following $\approx$ means asymptotical or approximate
equality and $\sim$ means proportionality), so the integral is infinite. Thus
the Bohr-Sommerfeld quantization is impossible.

What is the reason behind this?
The answer is that $\hat H$ is not bounded from below if $s < 0$. To see
this let us consider a square integrable wave function $\psi$ and define
    \begin{equation}
    \psi_\lambda (\delta) := \lambda \psi(\lambda\delta).
    \lbq{660}\end{equation}
The $\psi_\lambda$-s are normalized to 1; the expectation value of $\hat
H$ in a state $\psi_\lambda$ is
    \begin{equation}
    (\psi_\lambda, \hat{H} \psi_\lambda) = (\psi_\lambda, \hat
    H_1 \psi_\lambda) + (\psi_\lambda, \hat{H}_2 \psi_\lambda),
    \lbq{670}\end{equation}
where $\hat{H}_1$ and $\hat{H}_2$ denote the first and second term of $\hat
H$, respectively (cf. \rrf{540}). The scaling properties of $\hat{H}_1$
and $\hat{H}_2$ are such that
    \begin{equation}
    (\psi_\lambda, \hat{H}_1 \psi_\lambda) = (\psi, \hat{H}_1 \psi)
    \lbq{680}\end{equation}
and
    \begin{equation}
    (\psi_\lambda, \hat{H}_2 \psi_\lambda) = \lambda^2 (\psi, \hat{H}_2
    \psi) = \lambda^2 m s \int_{-\infty}^{\infty}
    \vert\psi(\delta)\vert^2 \frac{1}{\delta^2} \sqrt{2} \vert\delta\vert
    d\delta.
    \lbq{690}\end{equation}
In the r.h.s. of \rrf{690} $\lambda^2$ is multiplied by a negative number.
As a result if $\lambda$ increases to $\infty$ then $(\psi_\lambda, \hat
H \psi_\lambda)$ tends to $-\infty$.

This lack of a ground state causes the failure of the Bohr-Sommerfeld
quantization.
Fortunately the eigenvalues and eigenfunctions of $\hat H$ can be
determined exactly in all the cases $s > 0$, $s = 0$, $s < 0$. To do this we
have to solve the equation $ \hat H \psi = E \psi $
as a differential equation of second order. This equation has three singular
points: $\delta = \pm\infty$, which are regular singular points and
$\delta = 0$, which is an irregular sigular point. Consequently one has to
solve this equation restricted to the domains $\delta \in \R^+$ and $\delta
\in \R^-$ respectively and then to fit together the obtained
'half-eigenfunctions'. Because of the symmetry $\delta \leftrightarrow
-\delta$ it is enough to work on $\R^+$, the restriction of $\hat H$ to $\R^+$
will be denoted by $\hat H_+$.

In the case $s < 0$ the eigenvalue equation can be transformed to the Bessel
equation
    \begin{equation}
    z^2 w'' + z w' + (z^2 - \nu^2) w = 0
    \lbq{710}\end{equation}
by the substitutions
   \begin{equation}
    z = \frac{k}{\delta}
    \lbq{720}\end{equation}
and
    \begin{equation}
    \psi(\delta) = \frac{1}{\delta} \, w \! \left( \frac{k}{\delta}
    \right),
    \lbq{730}\end{equation}
where
    \begin{equation}
    \nu^2 = 1 - \frac{4mE}{\hbar^2},
    \lbq{740}\end{equation}
and
    \begin{equation}
    k = \sqrt{-\frac{4 m^2 s}{\hbar^2}}.
    \lbq{750}\end{equation}
The
two linearly independent solutions of \rrf{710}---existing for any complex
value of $\nu$---are $J_\nu(z)$ and $Y_\nu(z)$ (for the conventions and
properties concerning the Bessel functions cf. \cite{2}). Similarly, for $s >
0$ the transformations
    \begin{equation}
    z = \frac{\kappa}{\delta},
    \lbq{760}\end{equation}
    \begin{equation}
    \psi(\delta) = \frac{1}{\delta} \, w \! \left( \frac{\kappa}{\delta}
    \right)
    \lbq{770}\end{equation}
lead to the modified Bessel equation
    \begin{equation}
    z^2 w'' + z w' - (z^2 + \nu^2)w = 0
    \lbq{780}\end{equation}
with $\nu^2$ being the same as in \rrf{740} and
    \begin{equation}
    \kappa = \sqrt{\frac{4 m^2 s}{\hbar^2}}.
    \lbq{800}\end{equation}
Now the two solutions of \rrf{780} are the modified Bessel functions
$I_\nu(z)$ and $K_\nu(z)$. In the case $s = 0$ the transformation
\rrf{633} is
the most useful. The eigenfunctions of $\hat{H}_x$ are $\, \exp(\pm iKx)$,
where $K = \sqrt{2mE/\hbar^2 - 1/2} \,$ (cf. \rrf{640}). In the variable
$\delta$ they read
    \begin{equation}
    \frac{1}{\delta} e^{\pm i \sqrt{2} K \ln \delta}.
    \lbq{820}\end{equation}

In the cases $s > 0$, $s = 0 \,$ $\hat{H}_x$ is bounded from below. The
corresponding condition on the energy eigenvalues is $ E \geq
\frac{\hbar^2}{4m} $. For $s = 0$ it means that $K$ is a nonnegative real
number. In the case $s > 0$ the condition gives $\nu^2 \leq 0$ causing that
only the functions $I_{iu}(z)$ and $K_{iu}(z)$, $u \in \R$ mean energy
eigenfunctions. For $s < 0$ the Bessel functions with real indexes lead to $E
\leq \frac{\hbar^2}{4m}$ and the imaginary indexes correspond to $E >
\frac{\hbar^2}{4m}$.

To get more acquainted with the eigenfunctions let us carry out a simple check
of our physical picture that is based on $\hat{H}_x$. In the cases $s > 0$, $s
< 0$ the potential term of $\hat{H}_x$ decreases exponentially to zero as $x$
tends to $\infty$. Consequently we expect that for $x \rightarrow \infty$ the
eigenfunctions with $E \geq \frac{\hbar^2}{4m}$ behave as plane waves. (For $s
= 0$ this expectation is satisfied trivially.) To see whether this is the case
we will make use of the $z \approx 0$ behaviour of $J_{iu}(z)$ and
$I_{iu}(z)$. Using \rrf{2080} up to a constant of proportionality they are of
the form
    \begin{equation}
    J_{iu}(z) \approx I_{iu}(z) \sim z^{iu}.
    \lbq{822}\end{equation}
{}From \rrf{633} and \rrf{720} the connection between the variables $x$ and $z$
is
    \begin{equation}
    z = k e^{-\frac{x}{\sqrt{2}}}, \qquad x = \sqrt{2} \ln \! \left(
    \frac{k}{z} \right),
    \lbq{823}\end{equation}
(or, for $s > 0$, the similar formulas with $\kappa$ instead of $k$,) so we can
see that in the variable $x$ $J_{iu}$ and $I_{iu}$ are asymptotically plane
waves. The momentum corresponding to them is $ p = - \hbar u / \sqrt{2} $.
Considering that for $x \rightarrow \infty$ $V(x)$ tends not to zero but to
$\hbar^2/4m$ and quoting the connection between $ \nu = i u $ and $E$ we find
that the expectation 'kinetic energy $= p^2/2m$' is satisfied as well. The two
other eigenfunctions, $Y_{iu}$ and $K_{iu}$ are linear combinations of
$J_{iu}$ and $J_{-iu}$ resp. $I_{iu}$ and $I_{-iu}$. Thus they
also behave the way we expect from our physical picture.

    \section{Orthogonality and completeness}\label{orttel}

It will be important to form a complete orthogonal system from the
half-eigenfunctions, an orthogonal basis in $L^2(\R^+, \rho)$. In the case $s
= 0$ this is simple: the set $\{\exp(\pm iKx) \, \vert \, K \in [0,
\infty)\}$ is a complete orthogonal system (in the variable $x \in (-\infty,
\infty)$). Therefore the same can be said about the functions \rrf{820} in the
variable $\delta$ in $L^2(\R^+, \rho)$. For $s < 0$ it is shown in the
appendices that there exist several
independent choices of a complete orthogonal system. The different bases can
be indexed by a $p \in (0, 2]$, the corresponding eigenvectors (given in the
transformed form $w(z)$) are
    \begin{equation}
    \begin{array}{cc}
    J_q(z), & q = p, \, p + 2, \, p + 4, \, \ldots, \\ \\
    \cos \left( \frac{\pi}{2} p \right) J_0(z) + \sin \left( \frac{\pi}{2}
    p \right) Y_0(z), & \\ \\
    e^{-i \theta_p(u)} J_{iu}(z) + e^{i \theta_p(u)} J_{-iu}(z),
    & u \in (0, \infty),
    \end{array}
    \lbq{830}\end{equation}
where
    \begin{equation}
    e^{i \theta_p(u)} = \frac{\cos(\frac{\pi}{2} p) \sinh(\frac{\pi}{2} u)
    + i \sin(\frac{\pi}{2} p) \cosh(\frac{\pi}{2}
    u)}{\sqrt{\cos^2(\frac{\pi}{2} p) \sinh^2(\frac{\pi}{2} u) +
    \sin^2(\frac{\pi}{2} p) \cosh^2(\frac{\pi}{2} u)}}.
    \lbq{840}\end{equation}
For the case $s > 0$ the appendices prove that only one complete orthogonal
system can be built from the functions $I_{iu}(z)$ and $K_{iu}(z)$, namely,
the set
    \begin{equation}
    \{K_{iu}(z) \, \vert \: u \in [0, \infty)\}
    \lbq{843}\end{equation}

What makes the difference that in the cases $s = 0$ and $s > 0$ the eigenbasis
is unique while for $s < 0$ there are infinitely many complete orthogonal
systems? The answer is in the self-adjointness of $\hat H_+$. For this reason
we determine the deficiency index of $\hat H_+$. $\hat H_+$ is a differential
operator of second order with real coefficients and two singular points $
\delta = 0 $, $ \delta = \infty $. Its deficiency index is equal to the number
of its orthogonal square integrable eigenfunctions corresponding to a {\em
non-real} eigenvalue (cf. \cite{3}) (the deficiency index does not depend on
the eigenvalue chosen). In the case $s > 0$ the deficiency index is zero as
for a fixed non-real $\nu^2$ none of the two linearly independent
eigenfunctions---$I_\nu(z)$ and $K_\nu(z)$ in the variable $z$---is square
integrable (cf. Appendix A). In the case $s < 0$ $J_\nu$ is square integrable
while $Y_\nu$ is not (we can choose Re $\! \nu > 0$ without loss of
generality, see Appendix A again). Thus in this case the deficiency index is
one. For $s = 0$ the deficiency index is zero, which can be seen most easily
in the variable $x$.

Now, starting with the case $ s < 0 $,  we recall a theorem of \cite{3}, which
states that if the deficiency index is one then the operator has several
self-adjoint extensions. \cite{3} also gives a condition for the different
domains of definition of the different self-adjoint extensions. For
$\hat H_+$ this condition says that a function $\psi(\delta)$ lying in
the domain of definition of a self-adjoint extension has (to be smooth
enough, cf. \cite{3}, and) to satisfy
    \begin{equation}
    \lim_{\delta \rightarrow 0} \left[ \delta^3 \left( \psi^*
    \frac{dU_\nu^\vartheta}{d\delta} - \frac{d\psi^*}{d\delta} U_\nu^\vartheta
    \right) \right] = \lim_{\delta \rightarrow \infty} \left[ \delta^3
\left(
    \psi^* \frac{dU_\nu^\vartheta}{d\delta} - \frac{d\psi^*}{d\delta}
    U_\nu^\vartheta \right) \right],
    \lbq{846}\end{equation}
where
    \begin{equation}
    U_\nu^\vartheta (\delta) = \frac{1}{\delta} J_\nu \left(
\frac{k}{\delta}
    \right) + e^{i \vartheta} \frac{1}{\delta} J_\nu^* \left(
\frac{k}{\delta}
    \right)
    \lbq{850}\end{equation}
with a $\vartheta \in [0, 2 \pi)$ and a $ \nu \in \C \setminus \R $,
$\mbox{Re} \, \nu > 0$. $\vartheta$ and $\nu$ together index the different
self-adjoint extensions.

Then if one examines which eigenfunctions are included in the domain of
definition of a self-adjoint extension indexed by an arbitrarily chosen value
of $\vartheta$ and $\nu$, a straightforward if lengthy calculation shows that
these eigenfunctions are exactly the ones that form one of the complete
orthogonal systems \rrf{830}. The number $p$ which characterises this system
is expressed by $\vartheta$ and $\nu$ as
    \begin{equation}
    p = \mbox{Re} \, \nu + \frac{2}{\pi} \arcsin
    \left( \frac{\sinh(\frac{\pi}{2} \mbox{Im} \, \nu)
    \sin(\frac{\vartheta}{2})}{\sqrt{\sinh^2(\frac{\pi}{2}
    \mbox{Im} \, \nu) \sin^2(\frac{\vartheta}{2}) + \cosh^2(\frac{\pi}{2}
    \mbox{Im} \, \nu) \cos^2(\frac{\vartheta}{2})}} \right) \pmod{2}
    \lbq{860}\end{equation}
(For deriving this result one can make use of the asymptotics of the Bessel
functions and their derivatives, for the asymptotics cf. Appendix A).
Hence the multiplicity of the eigenbases origins in the multiple self-adjoint
extensions of the differential operator $\hat H_+$.

In the cases $s > 0$ and $s = 0$ the deficiency index is zero. The appropriate
theorem of \cite{3} states that then the operator is self-adjoint.
Consequently, the domain of definition is unique. All the eigenfunctions (or,
more precisely, all the wave packets superposed from the
eigenfunctions---remember that for both $s > 0$ and $s = 0$ all the
eigenfunctions are non-normalizable) are lying in the domain of definition so
the eigenbasis is unique as well (up to linear equivalence, i.e. except from
trivial phase factors or, in the case $s = 0$, choosing two linear
combinations of $\exp(iKx)$ and $\exp(-iKx)$ instead of $\exp(iKx)$ and
$\exp(-iKx)$).

    \section{The eigenfunctions on the whole configuration
    space}\label{egeszsfv}

To investigate the eigenfunctions of the full system the task is to sew
together the half-eigenfunctions and build up a complete orthogonal system of
'whole-eigenfunctions' (in the following: eigenfunctions). In usual quantum
mechanical systems, i.e. with a Hamiltonian of the form \rrf{550} and with no
weight function in the scalar product, the conditions for fitting parts of an
eigenfunction together are the continuity of the eigenfunction and the
continuity---or in special cases a given jump---of its (space) derivative. Now
we cannot expect that such conditions work. In fact, the $\delta \rightarrow
0$ behaviour of the half-eigenfunctions proves to be $\delta^{-\frac{1}{2}}
\cos(k/\delta + \mbox{const.})$ in the case $s < 0$, $\delta^{-1} \exp(\pm
i\sqrt{2} K \ln\delta)$ if $s = 0$ and $\delta^{-\frac{1}{2}}
\exp(-\kappa/\delta)$ if $s > 0$. Thus this kind of fitting together is
impossible. The situation is not better in the variable $x$ either, the
half-eigenfunctions tend to $0$ in the limit $x \rightarrow -\infty$ in the
cases $s < 0$ and $s > 0$, while for $s = 0$ they behave as $\exp(\pm iKx)$.
This infinite growth or decrease and infinitely rapid oscillating behaviour of
the half-eigenfunctions origins in the irregular singularity of the
Hamiltonian at $\delta = 0$.

Fortunately, the probability current is finite at $\delta \rightarrow 0$, it
is this quantity we are able to fit. However, in our case the probability
current is not of the usual form though. By deriving the continuity equation
for the probability density from the Schr\"odinger equation the probability
current proves to be
	\begin{equation}
	\frac{\hbar}{2 \sqrt{2} mi} |\delta|^3 \left( \psi^*
\frac{d \psi}{d \delta} - \frac{d \psi^*}{d \delta} \psi \right).
	\lbq{valaram}\end{equation}
Any $\psi$ can be expressed
as a linear combination of eigenfunctions $\varphi_k$, which makes it possible
to decompose the probability current as a sum of
	\begin{equation}
	\frac{\hbar}{2 \sqrt{2} mi}
	|\delta|^3 \left( \varphi_k^* \frac{d \varphi_l}{d \delta} - \frac{d
	\varphi_k^*}{d \delta}
	\varphi_l \right).
	\lbq{reszvalaram}\end{equation}
It can be verified that in each cases $s < 0$, $s = 0$,
$s > 0$
such a quantity has a well defined finite limit for $\delta \rightarrow 0$ so
the probability current is also finite at $\delta \rightarrow 0$.

We do not fit the probability current directly but carry out an equivalent
procedure. In fact, fitting the probability current of the half-eigenfunctions
is to ensure that the norm of a whole wave function does not change in time.
The latter is equvivalent to the self-adjointness of the whole Hamiltonian. We
know that the eigenfunctions of a self-adjoint Hamiltonian are orthogonal.
Conversely, a complete orthogonal system of the eigenfunctions of the
Hamiltonian as a differential operator defines a self-adjoint Hamiltonian from
the differential operator on an everywhere dense set in $L^2(\R, \rho)$, which
is our purpose. That's why it is enough---while it is more interesting as
well---to build up complete orthogonal systems out of the eigenfunctions
instead of fitting the probability current.

Let us start with the case $s < 0$. An eigenfunction $\Phi(\delta)$ is
generally of the form
    \begin{equation}
    \left\{ \begin{array}{cc}
    \alpha \, \varphi(-\delta) & \mbox{if $\delta < 0$,} \\
    \beta  \, \varphi( \delta) & \mbox{if $\delta > 0$,}
    \end{array} \right.
    \lbq{alfabeta}\end{equation}
where $\varphi$ is a half-eigenfunction, defined on $\R^+$. {}From this it
follows immediately that at most two linearly independent eigenfunctions can
correspond to an eigenvalue in a (whole-)eigenbasis. Another important
observation is that if a value $p$ corresponds to $\varphi$---the
index of the half-eigenbasis $\varphi$ is a member of---, this $p$
characterizes $\Phi$ as well. Now let us suppose that a complete orthogonal
system of
eigenfunctions does not include two linearly independent
eigenfunctions that correspond to the same eigenvalue {\it and} have the
same value $p$. (Later we will examine the other case as well, i.e. when one
can find
two such eigenfunctions in the system.) In this case there must
be at least one
eigenfunction in this eigenbasis with a different $p$. Otherwise we do not
have completeness: there exist functions that are orthogonal to any basis
vector but are not identically zero; such an example is a whole-eigenfunction
that is not included in the basis but has the same value $p$.

For two eigenfunctions having different $p$-s $(\; , \; )_+$ and $(\; , \;
)_-$, the restriction of their scalar product to the positive resp. negative
half of the configuration space are not zero, consequently they are orthogonal
only if one of them is of the form
    \begin{equation}
    \mbox{const.} \left\{ \begin{array}{cc}
    \varphi(-\delta)        & \mbox{if $\delta < 0$,} \\
    \lambda \, \varphi(\delta) & \mbox{if $\delta > 0$,}
    \end{array} \right.
    \lbq{1}\end{equation}
and the other is of the form
    \begin{equation}
    \mbox{const.} \left\{ \begin{array}{cc}
    -\lambda^* \, \varphi(-\delta) & \mbox{if $\delta < 0$,} \\
    \varphi(\delta)             & \mbox{if $\delta > 0$,}
    \end{array} \right.
    \lbq{2}\end{equation}
(with a different $\varphi$ but) with the same complex $\lambda$ from the set
$ \{ |\lambda| \leq 1 \mbox{, if } |\lambda| = 1 \mbox{ then arg} \lambda \in
[0,\pi) \} $. Let $p_1$ denote the value $p$ of the eigenfunction of the first
form and $p_2$ the $p$ of the other one. The other eigenfunctions with $p_1$
also must have the form \rrf{1} (with the same $\lambda$) and the other
eigenfunctions with $p_2$ also must have the form \rrf{2}, in order to be
orthogonal to these two eigenfunctions. These forms ensure that the further
eigenfunctions are orthogonal to each other as well. Orthogonality also
excludes the existence of any eigenfunctions in the eigenbasis having a $p$
other than $p_1$ or $p_2$.

To examine completeness first let us see whether an arbitrary function
$\psi_1$ from $L^2(\R, \rho)$ having the form \rrf{1} (where now $\varphi$ is
not a half-eigenfunction but an arbitrary half-function) can be spanned by
these eigenfunctions. It is easy to see that this requirement is equivalent to
that the restriction of the eigenfunctions with $p_1$ to $\R^-$ have to form a
complete half-eigenbasis (a $\psi_1$ is orthogonal to the eigenfunctions with
$p_2$, hence only the eigenfunctions with $p_1$ contribute to it). After a
similar treatment of the $\psi_2$-s of the form \rrf{2} we conclude that a
complete system must consist of each of the eigenfunctions with $p_1$ (that
have the form \rrf{1}) and each of the eigenfunctions with $p_2$. Then if any
$ \psi \in L^2(\R, \rho) $ can be given as a sum of a $\psi_1$ and a $\psi_2$
then completeness is reached. With the notation
    \begin{equation}
    \psi(\delta) = \left\{ \begin{array}{cc}
    \psi_-(-\delta), & \delta < 0, \\
    \psi_+( \delta), & \delta > 0
    \end{array} \right.
    \lbq{psi}\end{equation}
the sum of the functions
    \begin{equation}
    \psi_1(\delta) := \left\{ \begin{array}{cc}
    \frac{   1   }{1 + |\lambda|^2}
    \left( \psi_-(-\delta) + \lambda^* \psi_+(-\delta) \right), &
    \delta < 0, \\
    \frac{\lambda}{1 + |\lambda|^2}
    \left( \psi_-( \delta) + \lambda^* \psi_+( \delta) \right), &
    \delta > 0
    \end{array} \right.
    \lbq{psi1}\end{equation}
and
    \begin{equation}
    \psi_2(\delta) := \left\{ \begin{array}{cc}
    \frac{-\lambda^*}{1 + |\lambda|^2}
    \left( -\lambda \psi_-(-\delta) + \psi_+(-\delta) \right), &
    \delta < 0, \\
    \frac{    1     }{1 + |\lambda|^2}
    \left( -\lambda \psi_-( \delta) + \psi_+( \delta) \right), &
    \delta > 0
    \end{array} \right.
    \lbq{psi2}\end{equation}
is $\psi$, thus the completeness of the considered system of
eigenfunctions---which we shall denote by $(p_1, p_2, \lambda)$---is proven.
We remark that the above decomposition of $\psi$ is a generalization of the
decomposition of a function to a sum of an even and an odd function, which is
actually the case $\lambda = 1$.

Now let us turn to the other case, i.e. when the complete orthogonal
system of whole-eigenfunctions includes two linearly independent
eigenfunctions with a same value $p$ and corresponding to the same eigenvalue.
In this case the other eigenfunctions must be of this $p$ as well, otherwise
they cannot be orthogonal to both of these eigenfunctions. Furthermore, the
eigenvalues of the eigenbasis must be identical with the eigenvalues of the
half-eigenbasis $p$ and must be doubly degenerated: in the case of a simply
degenerated or missing eigenvalue any (other) eigenfunction corresponding to
this eigenvalue is orthogonal to each eigenfunction from the system, which is
in contradiction with completeness. The constants $\alpha$ and $\beta$ (see
\rrf{alfabeta}) for the eigenfunctions of the system can be arbitrary, the
only requirement is that for each eigenvalue the corresponding two
eigenfunctions be linearly independent. The concrete values of these
$\alpha$-s and $\beta$-s are not important, they only embody a choice of two
basis vectors in a two dimensional linear subspace. Remarkably, such an
eigenbasis is linearly equivalent to a one which---in the spirit of our
notation---can be denoted by $(p, p, \lambda)$ (the equivalence holds for an
arbitrary $\lambda$). Based on this observation one can prove completeness the
same way as for a system $(p_1, p_2, \lambda)$.

We see that in contrast to the eigenbases $(p_1, p_2, \lambda)$ considered
earlier, these latter eigenbases are characterized by a single number $p$.
Each of the different eigenbases $(p_1, p_2, \lambda)$ and $p$ means a
different self-adjoint extention of the Hamiltonian as a differential
operator.

For $ s > 0 $ and $ s = 0 $ the method to establish an eigenbasis is the same
as for an eigenbasis with a single $p$ in the case $ s < 0 $. The difference
is that now one starts with only one half-eigenbasis, consequently one arrives
at only one eigenbasis (up to linear equivalence). As a result in these
cases the self-adjoint Hamiltonian is unique.

That for $ s < 0 $ two different half-eigenbases are needed in general for
one eigenbasis may seem unusual. However, this situation is just an
analogue of the case of the operator $-\partial^2 \! /\partial x^2$ on the
interval $[-\pi, \pi]$. This simple example will help us to understand - at
least to some extent - what happens here.

The complete orthogonal system of the eigenvectors of $-\partial^2 \!
/\partial x^2$ corresponding to the conditions
    \begin{equation}
    \Phi(-\pi) = \Phi(\pi) = 0
    \lbq{915}\end{equation}
consists of the functions
    \begin{equation}
    \sin \left[ \frac{n}{2} (x + \pi) \right], \qquad n = 1, 2, \ldots
    \lbq{920}\end{equation}
These eigenfunctions are either even or odd functions, the even ones can be
written as
    \begin{equation}
    \sin \left( \frac{k}{2} x \right), \qquad k = 2, 4, \ldots,
    \lbq{930}\end{equation}
the odd ones are of the form
    \begin{equation}
    \cos \left( \frac{l}{2} x \right), \qquad l = 1, 3, \ldots
    \lbq{940}\end{equation}
Let us examine how one can get this eigenbasis by building it up from the
half-eigenbases of $-\parc^2 \! /\parc x^2$, the eigenbases of the operator
restricted to $[0, \pi]$ and $[-\pi, 0]$.

The eigenfunctions of $[0, \pi]$ corresponding to the conditions
    \begin{equation}
    \varphi(0) = \varphi(\pi) = 0
    \lbq{950}\end{equation}
are
    \begin{equation}
    \sin mx, \qquad m = 1, 2, \ldots
    \lbq{960}\end{equation}
We see that these functions can be the building blocks of the functions
\rrf{930}, via an antisymmetric (i.e. odd) extension from $[0, \pi]$ to
$[-\pi, \pi]$. On the other side, there is no way to build up the functions
\rrf{940} from them as well. The solution is that we have to consider another
half-eigenbasis, namely, the one corresponding to the conditions
    \begin{equation}
    \varphi'(0) = \varphi(\pi) = 0.
    \lbq{970}\end{equation}
The eigenfunctions satisfying \rrf{970} are
    \begin{equation}
    \cos jx, \qquad j = \frac{1}{2}, \frac{3}{2}, \ldots,
    \lbq{980}\end{equation}
and clearly a symmetric (i.e. even) extension of them to $[-\pi, \pi]$ yields
the functions \rrf{940}.

The operators $-\parc^2 \! /\parc x^2$ and $\hat{H}$ are common in that
both commute with the space reflection operator, and in that both have several
self-adjoint extensions. These are the aspects why the operator $-\parc^2 \!
/\parc x^2$ is a good example to understand $\hat{H}$. However, crucial
differences can be observed between the two operators. One of them is that the
proof that a Hamiltonian with space reflection symmetry must have even and odd
eigenfunctions does not hold for $\hat{H}$. The other,
more serious difference is that in the case of $-\parc^2 \! /\parc x^2$ there
exists a physical principle to choose one from the different self-adjoint
extensions. This is possible because the operator $-\parc^2 \! /\parc x^2$
together with the interval $[-\pi, \pi]$ arises in physics as the Hamiltonian
of the system characterized by the potential
    \begin{equation}
    V(x) = \left\{ \begin{array}{ll} 0 & \mbox{if $\vert x \vert < \pi$,} \\
                                \infty & \mbox{if $\vert x \vert > \pi$.}
    \end{array} \right.
    \lbq{990}\end{equation}
Then the requirements that an eigenfunction be continous and that it be zero
where $V(x) = \infty$ assign one of the possible boundary conditions that
characterize the different self-adjoint extensions (this distinguished
condition is just \rrf{915}). In this case there is a way to choose the
'physical' self-adjoint extension of the Hamiltonian. For $\hat H$ there
is no such principle, all the self-adjoint extensions prove to be equal. We
have to face the fact that there is no unique quantum mechanics corresponding
to the $s < 0$ classical system.

In spite of the non-usual form of the Hamiltonian and the presence of the
nontrivial weight function, the $\hat H_x$ form of the Hamiltonian enables
us to give the physical interpretation of the results to some extent.

In the case $s < 0$ we expect that the two half-configuration spaces are
in physical connection, the particle can cross the border $\delta = 0$.
Simple calculations show that this expectation is satisfied for the
self-adjoint extensions $ (p_1, p_2, \lambda) $, there is a probability
flow from one half to the other one. Consider for example a wave function
$ \psi = c_1 \psi_1 + c_2 \psi_2 $ where $\psi_1$ and $\psi_2$ are
eigenfunctions, one having $p_1$ and the other having $p_2$. Though $
\frac{d}{dt} (\psi, \psi) = 0 $, $ \; \frac{d}{dt} (\psi, \psi)_+ = -
\frac{d}{dt} (\psi, \psi)_- \neq 0 $ for generic $c_1$ and $c_2$. Another
transparent possibility to show the physical connectedness of the two
halves is that one can easily find examples for a solution of the (time
dependent) Schr\"odinger equation where the expectation value of the
coordinate operator $\hat \delta$ is oscillating in time between a
positive and a negative value. However, in the case of the self-adjoint
extensions $p$ the two halves behave as two closed, independent
subsystems. The reason is that for these eigenbases the restriction of the
eigenfunctions on a half configuration space is a half-eigenbasis, causing
that $\hat H$ decouples to two self-adjoint half-operators.

The cases $s > 0$ and $s = 0$ are similar to the $s < 0$, $p$ one. The
Hamiltonian is simply a pair of two self-adjoint half-Hamiltonians, the two
parts of the configuration space are physically independent. This result is in
accord with the naive pictures of the $s > 0$ and $s = 0$ systems based on
$\hat{H}_x$. For $s > 0$ we can think of an exponentially increasing and thus
infinitely wide potential wall separating the two half-worlds, no wonder that
we find no tunneling from one side to the other. The situation is similar to
the quantum mechanics of the system with the potential
    \begin{equation}
    V(x) = \left\{ \begin{array}{ll} 0 & \mbox{if $\vert x \vert > a$,} \\
                                \infty & \mbox{if $\vert x \vert < a$.}
    \end{array} \right.
    \lbq{1000}\end{equation}
where it is meaningful to speak about the quantum mechanics of the system on
the whole configuration space, yet there is no physical connection between the
two allowed parts. In the case $s = 0$ we have two free theories, both on an
infinitely large configuration space (understood in the variable $x$). We may
argue that under such circumstances a wave packet starting from one side (e.g.
the $\delta > 0$ one) cannot reach the other side in a finite time period. We
cannot say any stronger concerning interpretation: these are the limits we are
forced into.

    \section{Coordinate independence}\label{kooflen}

The reduced quantum system, as we saw, posesses several unusual
properties. It is natural to ask whether these features are only
artifacts, caused by the special coordinate system
which was used for the definition of the reduced system. Therefore it is
worth examining the possibility to define the system in a coordinate
independent way.

The Hamiltonian of the unconstrained quantum theory---a multiple of the
Laplacian of the manifold $SL(2,R)$---and the scalar product \rrf{420}
are in
fact coordinate invariant. Consequently the question reduces to whether the
constraints can be given a coordinate independent form. In Sect. 5 the
constraints were imposed through the canonical momentum operators. The
definition $ \hat p_k := \frac{\hbar}{i} ( \parc_k + \frac{1}{2} \parc_k \ln
\sqrt{-h} ) $ does not define a covariant quantity because $h$ is not a
coordinate invariant scalar. That's why it is recommended to impose the
constraints independently of the canonical momentum operators.

In the spirit of the Lie derivative, let us introduce the following
derivation operators
    \begin{equation}
    (L_A \Psi)(g) = \frac{d}{ds} \Psi(e^{As} g) \Bigg|_{s = 0}, \qquad (R_A
    \Psi)(g) = \frac{d}{ds} \Psi (g e^{As}) \Bigg|_{s = 0}
    \lbq{LRdef}\end{equation}
for any $A \in sl(2,\R)$. The definition of $L_A$ and $R_A$ does not need any
coordinate system. Nevertheless, if expressing them using the coordinates
$\delta$, $a,$ $c$ one finds that $L_{e_{12}} = \parc_a$ and $R_{e_{21}} =
\parc_c$. Thus we obtained a coordinate independent reformulation of the
constraints \rrf{500}.

One can feel the need for checking whether the operators $\frac{\hbar}{im}
L_{e_{12}}$ and $\frac{\hbar}{im} R_{e_{21}}$ are really the quantum
equivalents of the classical quantities $\mbox{Tr} \left[ e_{12} \, \dot g
g\inv \right]$ and $\mbox{Tr} \left[ e_{21} \, g\inv \dot g \right]$. The
following heuristic argument makes this relation visible.

We consider a wave packet which is in some sense the most similar to a
classical trajectory, determine the expectation value of $\frac{\hbar}{im}
L_A$, and compare it to $\mbox{Tr} [A \, \dot g \, g\inv]$ computed on the
classical
trajectory the wave packet is similar to (the relation will hold for any $A
\in sl(2,\R)$ in general; the analogous treatment for $R_A$ is
straightforward).

In the case of the free quantum mechanics on a three dimensional Euclidean
space the Gaussian wave packet
    \begin{equation}
    \mbox{const.} \int d^3 \k \, e^{ -\frac{1}{2\sigma^2} (\k - \k_0)^2 }
e^{ i
    \k (\x - \x_0) }
    \lbq{sikGauss}\end{equation} is in some sense the best wave mechanical
analogue of a classical trajectory. It is well-localized both in position
and in momentum---around the position $\x_0$ and the wave vector $\k_0$.
We wish to define the analogue of this wave function in the case of
$SL(2,\R)$. The differences in the two configuration spaces are that
$SL(2,\R)$ does not have a linear structure and the natural metric on it
is not positive definite. Nevertheless, we 'flatten' $SL(2,\R)$ around a
chosen point $g_0$---which we want as the 'centre' of the wave packet---by
characterizing a point $g$ with a Lie algebra element denoted by $X_g$
defined by the relation $g = e^{X_g} \, g_0$. This definition is correct,
i.e. $X_g$ exists and is unique, if $g$ is in a small enough neighbourhood
of $g_0$, and this is enough for our purposes because we want to define
only a fairly well localized wave packet around $g_0$. $X_g$ is the
analogue of $\x - \x_0$ in the Euclidean case. As far as the indefinite
metric is concerned: that a wave packet is localized is actually a
question involving topology rather than metric. That's why we will not
use the natural metric $ \{ h_{kl} \} $, or, more
precisely, the
corresponding Killing form $G$, to express that our wave packet is
localized. ($G$ is the metric tensor
at the unity element of $SL(2,\R)$, acting on the Lie
algebra of $SL(2,\R)$, $G(U,V) = \mbox{Tr} [UV]$ ). Instead,
we introduce an arbitrary positive definite inner product $\tilde G$ in
$sl(2,\R)$ and apply this $\tilde G$ in the Gaussian modulus of the
integrand in \rrf{sikGauss}. This procedure is similar to the case of the
Minkowski space, where for example differentiability or other analytic
properties are defined not by the use of the Lorentzian metric but by an
arbitrarily chosen positive definite inner product; it turns out that
these properties are independent of the choice of this artificial inner
product. In the phase of the integrand of \rrf{sikGauss} we keep the
Killing form.

All in all, let us work with the wave packet
    \begin{equation}
    \Psi_{g_0, K_0} (g) = \mbox{const.} \int d^3 K e^{ -\frac{1}{2\sigma^2}
    \tilde G(K - K_0, K - K_0) } \, e^{ i G(K, X_g) }.
    \lbq{gorbeGauss}\end{equation} The width $\sigma$ will not have to be
specified. This wave packet is by definition localized around $K_0$,
evaluating the integral shows that the result is a Gaussian in position
(also in the sense of $\tilde G$), localized around $g_0$ as expected.

The function $g \mapsto \exp iG(K,X_g)$ is approximately an eigenfunction of
$\frac{\hbar}{im} L_A$ with eigenvalue $\frac{\hbar}{m} G(K,A)$. By this we
mean that the eigenvalue equation holds exactly at $g = g_0$ and is satisfied
approximately at a neighbourhood of $g_0$. Consequently the expectation value
of $\frac{\hbar}{im} L_A$ is $\frac{\hbar}{m} G(K_0,A)$---approximately, i.e.
in the limit when the width of the wave packet in position tends to zero. Then
what is left is to show that $\frac{\hbar}{m} K_0 = \dot g g\inv$ for the
classical trajectory corresponding to \rrf{gorbeGauss}. To obtain this we make
use of the observation that $\triangle = (G\inv)^{ij} L_{I_i} L_{I_j} =
(G\inv)^{ij} R_{I_i} R_{I_j}$ where $\{I_i\}$ is an arbitrary basis in
$sl(2,\R)$. The most straightforward verification of this formula is to prove
it in coordinates. {}From this one can derive that $g \mapsto \exp iG(K,X_g)$
is approximately an eigenfunction of $\hat{\cal H} =
-\frac{\hbar^2}{2m}\triangle$
with eigenvalue $\frac{\hbar^2}{2m} G(K,K)$. With the aid of these pieces of
information we are able to tell the time propagation of the wave function
\rrf{gorbeGauss}. If considering \rrf{gorbeGauss} as the state function at $t
= 0$ then after a small time period the state function is approximately
    \begin{equation}
    \Psi_{g_0, K_0}(g, t) = \mbox{const.} \int d^3 K e^{ -\frac{1}{2\sigma^2}
    \tilde G(K - K_0, K - K_0) } e^{ i G(K, X_g) } e^{ -\frac{i}{\hbar} \left(
    \frac{\hbar^2}{2m} G(K,K) t \right) }.
    \lbq{gorbeGausst-kor}\end{equation}
\rrf{gorbeGausst-kor} is also a localized wave packet, but its centre
is not $g_0$ but a $g(t)$. This $g(t)$ plays the role of the classical
position of the masspoint at $t$. Using a stationary phase argument and taking
into account that the modulus of the integrand of \rrf{gorbeGausst-kor} is
concentrated around $K_0$ the centre of the wave packet \rrf{gorbeGausst-kor}
is at $X_{g(t)} = \frac{\hbar}{m} K_0 t$. Consequently, $\dot g g\inv =
\frac{\hbar}{m} K_0$, and this is what we wanted to show.

We saw that both the unconstrained system and the constraints are actually
coordinate independent. After defining a system in a coordinate independent
way, one can use concrete parametrizations to examine its properties. Turning
to the concrete situation: all the properties explored in the coordinates
$\delta$, $a$, $c$ are valid everywhere where $\delta \neq 0$. For example,
the wave functions are scalars so the infinite growth and infinitely rapid
oscillating of the eigenfunctions is a coordinate independent fact, since this
is the behaviour of the eigenfunctions not {\it at} but {\it around} the
invalid point $\delta = 0$.

    \section{Conclusions}\label{concl}

We investigated the properties of the point particle version of the reduced
$SL(2,\R)$ WZNW model both on the classical and the quantum level, for all the
possible values of the constraint parameters. We found that the quantum theory
exhibits an analogous behaviour to the classical one. The cases where the two
parts are disconnected classically lead to two independent systems on the
quantum level as well, and in the cases where the half-systems have a physical
connection, this connection can also be found in the quantum theory. The only
exception is that there is a possibility for a classically connected case to
be disconnected quantum mechanically. This is possible because not only one
quantum theory corresponds to a classically connected case. Several
self-adjoint extensions of the Hamiltonian exist, including special ones where
the two half-systems turn out to be independent.

Classical mechanically bounded motions exist, with arbitrary large
negative energies, in the connected cases. The disconnected cases do not
allow bounded motions and energy is bounded from below. These properties
are also reflected on the quantum level. It is remarkable that the quantum
theory is formally consistent irrespective of the values of the constraint
parameters, while in the connected cases it leads to systems with a
Hamiltonian not bounded from below (no matter which self-adjoint extension
is chosen). Recently a method was proposed to discuss quantum mechanical
systems that exhibit such a behaviour \cite{4}. The method implements the
concept of Wilson renormalization. It would be interesting to carry out
such an analysis for the system studied here. Nevertheless, the method of
\cite{4} means a kind of distortion of the system, which is not the
purpose here as here we are interested in the properties of the original
system for we want to obtain indications how the quantum theory of the
corresponding field theory behaves. In Sect. 3 we have found
classical space-independent configurations with arbitrary large negative
energy in the connected cases. This and the quantum properties of the
masspoint version make it quite possible that the energy is essentially
not bounded from below in the quantum field theory.

That the energy is not bounded from below is not the only nontrivial property
of the connected case. The most striking result of our analysis is the
existence of several self-adjoint extensions corresponding to one classical
system. There is no -- physical or mathematical -- principle to choose one out
of them as the 'real' one. The origin of this behaviour is the strong
singularity at the border which separates the two half-systems. This
singularity is not present on the unconstrained level, it is a consequence of
the charasteristics of the constraints. As this singularity can also be
observed in the classical reduced field theory version \cite{1}, we expect to
face the problem of the non-unique self-adjoint energy operator on the quantum
level, i.e. in the quantum field theory of the reduced $SL(2,\R)$ WZNW model
as well.

The method applied here to present the quantum mechanics of the reduced system
was canonical quantization (supplemented by a coordinate independent
approach). Because of the nontrivial properties found it would be interesting
to examine this system by using other tools, geometric quantization or
functional integration, and see how these methods give account of the
characteristics of the theory.

Additionally we remark that recently a paper carried out an analysis of the
relativistic quantum mechanics of a free particle on the $SL(2,\R)$ manifold
\cite{RQM}. The problem studied there is independent from the one presented
here. Clearly, in \cite{RQM} the group $SL(2,\R)$ plays the role of the
(curved) {\it space-time} the particle exists in while in our case
$SL(2,\R)$
is the (configuration) {\it space} of the unconstrained system.

\bigskip
\bigskip

\begin{center} {\bf Acknowledgements} \end{center}

\bigskip

The author wishes to thank L. Palla, L. Feh\'er and Z. Horv\'ath for the many
useful discussions.

This work is supported by the Hungarian Research Fund, OTKA F 4011.

\appendix

    \section{Orthogonality}

Let us first collect the properties of the Bessel functions and modified Bessel
functions we will need in the following. $J_\nu(z)$ and $Y_\nu(z)$ are analytic
functions of $z$ on the whole complex plane except the negative real half line.
For a fixed $z$ ($z \neq 0$) both are integer functions of $\nu$ for any $\nu
\in \C$. $J_\nu(z)$ and $J_{-\nu}(z)$ are linearly independent for any value of
$\nu$ except $\nu = n$ (let $n$ denote an integer subsequently), when
    \begin{equation}
    J_{-n}(z) = (-1)^n J_n(z).
    \lbq{2010}\end{equation}
$J_\nu(z)$ and $Y_\nu(z)$ are always linearly independent. The following
formula
    \begin{equation}
    Y_\nu(z) = \frac{\cos(\pi\nu) J_\nu(z) - J_{-\nu}(z)}{\sin(\pi\nu)}
    \lbq{2020}\end{equation}
establishes a connection between the $J$-s and the $Y$-s; for $\nu = n$
\rrf{2020} is understood as a limit when $\nu \rightarrow n$.

For the modified Bessel functions $I_\nu(z)$ and $K_\nu(z)$ all the statements
mentioned above apply if we replace $J_\nu(z)$ by $I_\nu(z)$ and $Y_\nu(z)$ by
$K_\nu(z)$, with the exceptions that $I_{-n}(z) = I_n(z)$ and that
    \begin{equation}
    K_\nu(z) = \frac{\pi}{2} \, \frac{I_{-\nu}(z) - I_\nu(z)}{\sin(\pi\nu)}.
    \lbq{2050}\end{equation}
Under complex conjugation
    \begin{equation}
    J_\nu(z)^* = J_{\nu^*}(z)
    \lbq{2060}\end{equation}
is valid if $z$ is real. This property holds for $Y_\nu(z)$, $I_\nu(z)$ and
$K_\nu(z)$ as well.

The modified Bessel functions can be expressed in terms of the ordinary
Bessel functions. We will make use of the identity
    \begin{equation}
    I_\nu(z) = e^{-i \frac{\pi}{2} \nu} J_\nu(e^{i \frac{\pi}{2}} z) \qquad
    \left(-\pi < \mbox{arg} \, z \leq \frac{\pi}{2}\right).
    \lbq{2075}\end{equation}

The asymptotic behaviour of these four functions at $z \approx 0$ is
    \begin{equation}
    J_\nu(z) \approx I_\nu(z) \approx \frac{1}{\Gamma(1+\nu)}
    {\left( \frac{z}{2} \right)}^\nu,
    \qquad (\nu \in \C \setminus \{-1, -2, -3, \ldots\}),
    \lbq{2080}\end{equation}
    \begin{equation}
    Y_\nu(z) \approx - \frac{2}{\pi} K_\nu(z) \approx -\frac{1}{\pi}
    \Gamma(\nu) {\left( \frac{z}{2} \right) }^{-\nu}, \qquad (\mbox{Re} \,
    \nu > 0)
    \lbq{2090}\end{equation}
and
    \begin{equation}
    Y_0(z) \approx -\frac{2}{\pi} K_0(z) \approx \frac{2}{\pi} \ln z.
    \lbq{2100}\end{equation}
For $\vert z \vert \rightarrow \infty$ the ordinary and modified Bessel
functions behave as
    \begin{equation}
    J_\nu(z) \approx {Y_\nu}'(z) \approx \sqrt{\frac{2}{\pi z}} \cos \left[ z -
    \frac{\pi}{2} \left( \nu + \frac{1}{2} \right) \right] ,
    \qquad (\vert \mbox{arg} \, z
    \vert < \pi),
    \lbq{2110}\end{equation}
    \begin{equation}
    {J_\nu}'(z) \approx -Y_\nu(z) \approx -\sqrt{\frac{2}{\pi z}} \sin \left[ z
-
    \frac{\pi}{2} \left( \nu + \frac{1}{2} \right) \right],
    \qquad (\vert \mbox{arg} \, z
    \vert < \pi),
    \lbq{2120}\end{equation}
    \begin{equation}
    I_\nu(z) \approx {I_\nu}'(z) \approx \frac{e^z}{\sqrt{2 \pi z}} \qquad
    \left( \vert \mbox{arg} \, z \vert < \frac{\pi}{2} \right)
    \lbq{2130}\end{equation}
and
    \begin{equation}
    K_\nu(z) \approx -{K_\nu}'(z) \approx \sqrt{\frac{\pi}{2z}} e^{-z}
    \qquad \left( \vert \mbox{arg} \, z \vert < \frac{3 \pi}{2} \right).
    \lbq{2140}\end{equation}

Expression \rrf{2080} is valid not only for fixed $\nu$ and $z \approx 0$ but
holds also if $z$ is fixed at an arbitrary (not necessarily small) value and
$\vert \nu \vert \rightarrow \infty$, $\vert \mbox{arg} \, \nu \vert < \pi$. In
this case it can be combined with the Stirling formula giving the $\vert \nu
\vert \rightarrow \infty$ asymptotics of the gamma function resulting
    \begin{equation}
    J_\nu(z) \approx I_\nu(z) \approx
    \frac{1}{\sqrt{2\pi\nu}} {\left( \frac{ez}{2 \nu} \right)}^{\nu}.
    \lbq{2145}\end{equation}

After these necessary pieces of information let us start finding the possible
orthogonal systems of the eigenvectors of the case $s < 0$. The scalar product
of $\psi_1(\delta)$ and $\psi_2(\delta)$ in $L^2(\R^+, \rho)$ is
    \begin{equation}
    (\psi_1, \psi_2)_+ = \int_0^\infty \psi_1(\delta)^*
    \psi_2(\delta) \sqrt{2} \delta \, d\delta,
    \lbq{2150}\end{equation}
here the notation $(\; , \;)_+$ reminds us that this scalar product is taken in
$L^2(\R^+, \rho)$, i.e. on the positive half of the configuration space only.
Instead of the variable $\delta$ it will be more suitable to work in $z$. Under
the transformations \rrf{720} and \rrf{730} the integral \rrf{2150} transforms
to
    \begin{equation}
    \sqrt{2} \int_0^\infty w_1(z)^* w_2(z) \frac{dz}{z}.
    \lbq{2160}\end{equation}
We will study this integral by considering it between finite $a$ and $b$ and
then take the limit $a \rightarrow 0$, $b \rightarrow \infty$.

If $w_1(z)$ and $w_2(z)$ are eigenfunctions of the Bessel equation \rrf{710}
with indexes $\mu$ and $\nu$ respectively then such an integral can be easily
evaluated due to a formula of \cite{6}. This formula states that
    \begin{equation}
    \int_a^b A_\mu (z) B_\nu (z) \frac{dz}{z} = \frac{1}{\nu^2-\mu^2}
    { \left[ z \left( A_\mu (z) B_\nu'(z) - A_\mu'(z) B_\nu (z)
    \right) \right] }_a^b
    \lbq{2165}\end{equation}
($A, B = J \mbox{ or } Y,\: 0 < a \leq b < \infty$). Applying it together
with \rrf{2060} the integral \rrf{2160} between the limits $a$ and $b$ is
    \begin{equation}
    \frac{\sqrt{2}}{\nu^2-\mu^{*2}} {\left[ z \left( w_1(z)^* w_2'(z) -
    w_1'(z)^* w_2(z) \right) \right]}_a^b.
    \lbq{2170}\end{equation}

With the aid of \rrf{2170} let us determine the scalar product $ (J_\mu,
J_\nu)_+ $ ($ \mu, \nu \in \C $).
Substituting them into \rrf{2170} and using the asymptotic formulas
\rrf{2110} and \rrf{2120}, the contribution from the 'upper limit terms' (the
terms depending on $b$) in the limit $b \rightarrow \infty$ is
    \begin{equation}
    \frac{2\sqrt{2}}{\pi(\nu^2-\mu^{*2})} \sin \left[ \frac{\pi}{2} (\nu
    - \mu^*) \right].
    \lbq{2180}\end{equation}
On the other hand, from \rrf{2080} one can see that the 'lower limit terms'
behave as a $(\mu^* + \nu)$-th power of $a$ when $a \rightarrow 0$. If Re$\,
(\mu^* + \nu) > 0$ then the 'lower limit terms' tend to zero; so in this case
    \begin{equation}
    (J_\mu, J_\nu)_+ = \frac{2\sqrt{2}}{\pi(\nu^2-\mu^{*2})} \sin \left[
    \frac{\pi}{2} (\nu - \mu^*) \right] .
    \lbq{2190}\end{equation}
If Re$\, (\mu^* + \nu) < 0$ then the integral diverges for $a \rightarrow 0$.
If Re$\, \mu > 0$ then the $\nu \rightarrow \mu$ limit gives the norm of
$J_\mu$:
    \begin{equation}
    (J_\mu, J_\mu)_+ = \frac{1}{\sqrt{2} \, \mbox{Re} \, \mu}.
    \lbq{2200}\end{equation}
Similarly we find that $(J_\mu, Y_\nu)_+$ is finite if Re$\, (\mu^* + \nu) >
0$ and infinite if Re$\, (\mu^* + \nu) < 0$, while the scalar product $(Y_\mu,
Y_\nu)_+$ always diverges. As a special case of this latter we also see that
the $Y_\mu$-s are non-normalizable, for any complex value of $\mu$.

We want to build complete orthogonal systems out of the $J_\mu$-s and
$Y_\mu$-s, where now only real and imaginary $\mu$-s are allowed (cf.
Sect. 6). The role of the Bessel functions with real respectively
imaginary index
is to span the wave functions that have an energy expectation value $ E >
\frac{\hbar^2}{4m} $ resp. $ E \leq \frac{\hbar^2}{4m} $. With a real $\mu$
only the $ J_{\mu \geq 0} $-s and $Y_0$ can be taken into account. In fact, a
$J_{\mu < 0}$ or an $Y_{\mu \neq 0}$ cannot be orthogonal to a Bessel function
with imaginary index as their scalar product diverges. {}From \rrf{2190} it
follows that a maximal, pairwise orthogonal set of $J_{\mu > 0}$-s is $ \{
J_p, J_{p+2}, J_{p+4}, \ldots \} $, where $ p \in (0, 2] $. With a $p$ given,
the only linear combination of a $J_{iu}(z)$ and $J_{-iu}(z)$ that is
orthogonal to $J_p(z)$ is the one given in \rrf{830}, as one finds with the
aid of \rrf{2190}. This linear combination proves to be orthogonal to the
eigenfunctions $J_{p+2}$, $J_{p+4}$, $\ldots$ as well. In the end, there
exists one linear combination of $J_0(z)$ and $Y_0(z)$ that is orthogonal to
$J_p$ (and, as turns out, to the functions $J_{p+2}$, $J_{p+4}$, $\ldots$ and
$ \exp [ -i \theta_p (u) ] J_{iu} (z) + \exp [ i \theta_p (u) ] J_{-iu} (z) $,
$ u \in (0, \infty) $ as well). This linear combination can be obtained either
by using the scalar product or as the $u \rightarrow 0$ limit of the
eigenfunctions $ \exp [ -i \theta_p (u) ] J_{iu} (z) + \exp [ i \theta_p (u) ]
J_{-iu} (z) $.

What is left to check is the mutual orthogonality of the eigenfunctions with
imaginary index. We will prove that
    \begin{equation}
    {\left( c(u) \left[ e^{-i\theta_p(u)} J_{iu} + e^{i\theta_p(u)}
    J_{-iu} \right], c(v) \left[ e^{-i\theta_p(v)} J_{iv} +
    e^{i\theta_p(v)} J_{-iv} \right] \right)}_+ = \delta(u - v)
    \lbq{2210}\end{equation}
where
    \begin{equation}
    c(u) = \frac{\vert \Gamma(1+iu) \vert}{\sqrt{2\sqrt{2} \pi}} =
    \sqrt{\frac{1}{2\sqrt{2}} \frac{u}{\sinh(\pi u)}}
    \lbq{2220}\end{equation}
(for the properties of the gamma function see for example
\cite{2}). As $u$ and $v$ run over the positive real numbers only and not on
the whole real line, the Dirac delta distribution must be understood here to
act on the test functions which are defined on the positive half of the real
line and are smooth functions of compact support vanishing at the origin. In
the following $u$ will be treated as a variable---i.e. the variable of the
test functions and the kernel functions---and $v$ as a fixed parameter.

To inspect the scalar product let us consider \rrf{2170} in our case. A
simple calculation involving the use of \rrf{2180} and some trigonometrical
identities shows that the contribution of the upper limit tends to $0$ when
$b \rightarrow \infty$. The lower limit terms are asymptotically
    \begin{equation}
    -\frac{\sqrt{2} c(u)c(v)}{i} \left[ \frac{e^{-i \left( \theta_p(u) -
    \theta_p(v) \right) }}{(u - v) \Gamma(1+iu) \Gamma(1-iv)} {\left(
    \frac{a}{2} \right)}^{i(u - v)} + \ldots \: \right]
    \lbq{2230}\end{equation}
where the \ldots stands for three other terms which can be obtained from the
first one by the substitutions $u \rightarrow -u$, $v \rightarrow -v$, or both,
respectively. Introducing $\Lambda = \ln \frac{2}{a}$, two of these terms have
the form $f(u, v) \sin \Lambda(u+v) + g(u, v) \cos \Lambda(u+v)$, where $f$ and
$g$ are smooth functions of both $u$ and $v$. It is a well-known fact that the
regular distributions $ \sin \Lambda x$ and $\cos \Lambda x$ tend to zero if
$\Lambda \rightarrow \infty$. This property does not change if we multiply them
by a smooth function, thus these terms give zero in the limit $\Lambda
\rightarrow \infty$ $(a \rightarrow 0)$.

With \rrf{2220} the two other terms can be written in the following way
    \[- \frac{1}{2 \pi i (u-v)} \left[ \left( e^{-i[ \theta_p(u) -
    \theta_p(v) + \mbox{arg} \, \Gamma(1+iu) - \mbox{arg} \, \Gamma(1+iv)]}
    - 1 \right) {\left( \frac{a}{2} \right)}^{i(u-v)} \right. \]
    \begin{equation}
    \left. + \:
    {\left( \frac{a}{2} \right)}^{i(u-v)} \right] - (u \leftrightarrow v).
    \lbq{2240}\end{equation}
{}From the properties of $\theta_p$ and the gamma function it follows that
the function $[\exp(\ldots) - 1] / (u-v)$ behaves smoothly even if $u
\rightarrow v$. Consequently the distributions coming from the first and the
third terms also tend to zero. What remained is equal simply to $\sin[\Lambda
(u-v)] \, / \pi(u-v)$. It is well-known that $\sin \Lambda x / \pi x
\rightarrow \delta(x)$ in the limit $\Lambda \rightarrow \infty$, so \rrf{2210}
is proven.

We close Appendix A by showing that in the case $s > 0 \, $ $\{K_{iu}(z) \,
\vert \: u \in [0, \infty)\}$ is the only possible orthogonal system built
from the eigenfunctions. For this let us consider the scalar product
    \begin{equation}
    {\left( c_+(u) I_{iu} + c_-(u) I_{-iu} , c_+(v) I_{iv} + c_-(v) I_{-iv}
    \right)}_+
    \lbq{2260}\end{equation}
(here $u, v > 0$ again). In the case $s < 0$ we had formula \rrf{2165} to
evaluate the integral corresponding to this scalar product. Repeating the proof
of \rrf{2165} given in \cite{6} one can obtain a corresponding result in the
case of the modified Bessel functions. The formula one gets turns out to be
exactly of the form of \rrf{2165} with $A$ and $B$ denoting now $I$ or $K$.
Thus we can study the scalar product similarly as we did in the case $s < 0$.
Let us evaluate \rrf{2260} with the aid of
\rrf{2165}, the 'upper limit terms' give asymptotically
    \begin{eqnarray}
    \frac{\sqrt{2}}{i\pi(u^2-v^2)} \left[ c_+(u)^* c_+(v) (e^{-\pi u}  -
    e^{\pi v}) + c_+(u)^* c_-(v) (e^{-\pi u}  - e^{-\pi v}) \right. \nonumber
    \\ \left. \mbox{} + c_-(u)^* c_+(v) (e^{\pi u} - e^{\pi v}) + c_-(u)^*
    c_-(v) (e^{\pi u} - e^{-\pi v}) \right].
    \lbq{2270}\end{eqnarray}
Using the asymptotics \rrf{2080} we can see that
the 'lower limit terms' behave as $(a/2)^{i(\pm u \pm v)}$, just like in the
case $s < 0$. Two of them vanishes if $a \rightarrow 0$. After a trick similar
to \rrf{2240} the nonzero contribution of the two other terms is a sum of a
$\sin \Lambda(u-v) / (u-v)$ and a $\cos \Lambda(u-v) / (u-v)$ term. In the
limit $\Lambda \rightarrow \infty$ the first of them leads to a Dirac delta.
The second one is not, consequently its coefficient must be equal to zero.
This gives the condition
    \begin{equation}
    c_+(u)^* \: c_+(v) = c_-(u)^* \: c_-(v).
    \lbq{2280}\end{equation}
On the other hand, \rrf{2270} is a smooth function for $u \neq v$.
Orthogonality requires that \rrf{2270} must be equal to zero for any $u \neq
v$. These two conditions together yield
    \begin{equation}
    \left[ c_+(u)^* + c_-(u)^*\right] \left\{ c_+(v) e^{\pi u} + c_-(v)
    e^{-\pi u} - c_+(v) e^{\pi v} - c_-(v) e^{-\pi v} \right\} = 0.
    \lbq{2290}\end{equation}
\rrf{2290} holds for any $u, v$, $u \neq v$ only if $c_+(u) + c_-(u) = 0$,
which we wanted to prove (cf. \rrf{2050}).

We note that by using \rrf{2165} and the asymptotics \rrf{2080}--\rrf{2140}
$I_\nu(z)$ and $K_\nu(z)$ prove to be not square integrable for any complex
value of $\nu$. Another remark is that the eigenfunctions in \rrf{830} and
\rrf{843} are real (cf. \rrf{2060}).

    \section{Completeness}

Here we prove the completeness of the orthogonal systems \rrf{830} and
\rrf{843}. Formulating the completeness of a system \rrf{830} in the variable
$x$ reads
    \begin{equation}
    S(x_1, x_2) + I(x_1, x_2) = \delta (x_1 - x_2)
    \lbq{2300}\end{equation}
with
    \begin{equation}
    S(x_1, x_2) = \sum_{q \in p + 2\Z^+} \sqrt{2} \, q \, J_q(z_1) J_q(z_2)
    \lbq{2310}\end{equation}
and
    \begin{eqnarray}
    I(x_1, x_2) = \int_0^\infty \frac{du \: u}{2\sqrt{2} \sinh \pi u} \left[
    e^{-i\theta_p(u)} J_{iu}(z_1) + e^{i\theta_p(u)} J_{-iu}(z_1) \right]
    \times \nonumber \\ \mbox{} \left[ e^{-i\theta_p(u)} J_{iu}(z_2)
    + e^{i\theta_p(u)} J_{-iu}(z_2) \right],
    \lbq{2320}\end{eqnarray}
where
    \begin{equation}
    z_1 = k e^{-\frac{x_1}{\sqrt{2}}}, \qquad z_2 = k
    e^{-\frac{x_2}{\sqrt{2}}}.
    \lbq{2325}\end{equation}
The convergence of the infinite sum \rrf{2310} is guaranteed by the
asymptotics \rrf{2145}.

First let us prove \rrf{2300} in the case when $p = 2$. We consider the
integral \rrf{2320} between $0$ and $\Lambda$, $\Lambda = N + 1/2$, $N \in
\Z^+$. The substitution $\nu := iu$ transforms the integral to
    \begin{equation}
    \int_0^{i\Lambda} \frac{d\nu \: \nu}{i 2\sqrt{2} \sin \pi\nu} \left[
    J_\nu(z_1) +
    J_{-\nu}(z_1) \right] \left[ J_\nu(z_2) + J_{-\nu}(z_2) \right].
    \lbq{2330}\end{equation}
As the integrand of \rrf{2330} is invariant under the transformation $\nu
\rightarrow -\nu$ \rrf{2330} can be written as
    \begin{equation}
    \frac{1}{2} \int_{-i\Lambda}^{i\Lambda} \frac{d\nu \:
    \nu}{i 2\sqrt{2} \sin
    \pi\nu} \left[ J_\nu(z_1) + J_{-\nu}(z_1) \right] \left[ J_\nu(z_2)
    + J_{-\nu}(z_2) \right].
    \lbq{2340}\end{equation}
Now we change the contour of this integral to a half circle (denoted by $C$)
starting from the point $-i\Lambda$, running in the half plane $\mbox{Re} \,
\nu > 0$ of the complex $\nu$-plane and ending at $i\Lambda$. The difference of
\rrf{2340} and this new integral can be expressed by the residues of the poles
of the integrand lying in the region bordered by the two contours. As the $J$-s
behave analytically, poles arise only from $\sin \pi\nu$, at the values $\nu =
1, 2, 3, \ldots N$ (at $\nu = 0 \,$ $\nu / \sin(\pi\nu)$ is not singular).
The contribution of the residues is
    \begin{equation}
    - \frac{2 \pi i}{i 4 \sqrt{2}} \sum_{n = 1}^N \mbox{Res}_n
    \lbq{2350}\end{equation}
where, by using \rrf{2010}, $\mbox{Res}_n$ turns out to be equal to $\left(
\frac{4n}{\pi} \right) J_n(z_1) J_n(z_2)$ if $n$ is even and is zero if $n$ is
odd. We can see that \rrf{2350} is just the opposite of \rrf{2310} in the limit
$N \rightarrow \infty$ so from now on we have to prove that
    \[
    \int_C
    \frac{d\nu \: \nu}{i 4\sqrt{2} \sin \pi\nu} \left[ J_\nu(z_1) +
    J_{-\nu}(z_1)
    \right] \left[ J_\nu(z_2) + J_{-\nu}(z_2) \right] = \; \; \; \; \; \;
    \;\;\;\;\;\;\;\;\;\;\;\;\;\;\;\;\;\;\;\;\;\;\;\;\;\]
    \begin{equation}
    \; \; \; \; \; \; \; \;\;\;\;
    \int_C \frac{d\nu \: \nu}{i
    4\sqrt{2} \sin \pi\nu}J_\nu(z_1) J_\nu(z_2) + \int_C \frac{d\nu \:
    \nu}{i 4\sqrt{2}
    \sin \pi\nu} J_{-\nu}(z_1) J_{-\nu}(z_2) +
    \lbq{2359}\end{equation}
    \[ \int_C
    \frac{d\nu \:
    \nu}{i 4\sqrt{2} \sin \pi\nu} \left[ J_\nu(z_1) J_{-\nu}(z_2) +
    J_{-\nu}(z_1) J_\nu(z_2) \right]
    \;\;\;\;\;\;\;\;\;\;\;\;\;\;\;\;\;\;\;\;\;\]
tends to $\delta(x_1 - x_2)$ when $\Lambda \rightarrow \infty$.

We will perform the proof in three steps. We start by showing that the first
and the second terms of the r.h.s. of \rrf{2359} are equal. The second step
proves that these terms are zero in the limit $\Lambda \rightarrow \infty.$
Thirdly it is shown that the third term tends to $\delta(x_1 - x_2)$ as
$\Lambda \rightarrow \infty$.

To see that the first and the second terms are equal let us make the
substitution $\nu \rightarrow -\nu$ in the second term. The integrand of the
resulting integral is the same as in the first term (up to a factor of $-1$),
but the contour $C'$ is a half circle starting from $i\Lambda$, running through
the half plane $\mbox{Re} \, \nu < 0$
and arriving at $-i\Lambda$. Let us change
the direction of $C'$, this causes another factor of $-1$ in the integral. Now
we can change this contour $C''$ to $C$, thus we arrive at the first term, plus
the contribution of the residues coming from the poles lying between $C''$ and
$C$. Fortunately this contribution
    \begin{equation}
    \sum_{n = -N}^N \mbox{Res}_n
    \lbq{2370}\end{equation}
is zero because, as a consequence of \rrf{2010}, $\mbox{Res}_n = -
\mbox{Res}_{-n}$.

In the second step we are interested in the $\Lambda \rightarrow \infty$
behaviour of the integral
    \begin{equation}
    \int_C \frac{d\nu \: \nu}{\sin \pi\nu} J_\nu(z_1) J_\nu(z_2).
    \lbq{2380}\end{equation}
Writing $\nu$ in the form of $\Lambda \exp(i \varphi)$ this integral can be
expressed as
    \begin{equation}
    \int_{-\frac{\pi}{2}}^{\frac{\pi}{2}} d\varphi \frac{i \Lambda^2 e^{2i
    \varphi}}{\sin(\pi\Lambda e^{i \varphi})} J_{\Lambda e^{i \varphi}}(z_1)
    J_{\Lambda e^{i \varphi}}(z_2).
    \lbq{2390}\end{equation}
It will be enough if we show that the integral of the integrand's modulus
tends to zero.

We will use asymptotic expressions for analysing the $\Lambda \rightarrow
\infty$ behaviour. By doing so some care will be needed. That's why we divide
the domain of integration $[-\frac{\pi}{2}, \frac{\pi}{2}]$ into three parts:
$A := [-\frac{\pi}{2}, -\frac{\pi}{4}]$, $B :=
[-\frac{\pi}{4}, \frac{\pi}{4}]$
and $C := [\frac{\pi}{4}, \frac{\pi}{2}]$. It can be proved easily that for a
$\varphi \neq 0$ and a large enough $\Lambda$
    \begin{equation}
    \vert \sin[\pi\Lambda e^{i \varphi}] \vert \approx \frac{1}{2}
    e^{\pi\Lambda \sin \vert \varphi \vert }.
    \lbq{2400}\end{equation}
Let us apply
\rrf{2400} in the domains $A$ and $C$. Combining it with \rrf{2145}
the modulus of the integrand of \rrf{2390} is asymptotically equal to
    \begin{equation}
    \frac{\Lambda}{\pi} \, e^{ -2\Lambda [(\ln \Lambda - \zeta) \cos \varphi +
    (\frac{\pi}{2} - \vert \varphi \vert) \sin \vert \varphi \vert]},
    \lbq{2410}\end{equation}
where $\zeta = \frac{1}{2} \ln(z_1 z_2 / 4) + 1$ is a quantity independent of
$\varphi$ and $\Lambda$. We can see that the integral of \rrf{2410} on $A$ or
$C$ gives
the same result. Thus we will consider this integral only on $C$, for
example.

If $\varphi \in [\pi/4, \pi/2]$ then the inequalities $\, \cos \varphi \geq 1
- (2/\pi) \varphi \,$ and $\, (\pi/2 - \varphi) \sin \varphi \geq 0$ hold,
helping us to give an upper estimate of \rrf{2410}
    \begin{equation}
    \frac{\Lambda}{\pi}
    \, e^{-2\Lambda (\ln \Lambda - \zeta) (1 - \frac{2}{\pi}
    \varphi)}
    \lbq{2420}\end{equation}
if $\ln \Lambda > \zeta$. The integral of \rrf{2420} on $C$ can be calculated
easily. The result is less then $[4(\ln \Lambda - \zeta)]^{-1}$, which is a
quantity tending to zero if $\Lambda \rightarrow \infty$.

In domain $B$ we cannot use \rrf{2400} but here it is enough to work with the
inequality
    \begin{equation}
    \vert \sin[\pi\Lambda e^{i \varphi}] \vert \geq 1.
    \lbq{2430}\end{equation}
For proving \rrf{2430} it is not hard to show that $\vert \sin[\pi\Lambda
\exp(i \varphi)] \vert$ takes its minimum in $\varphi = 0$ as $\varphi$ varies
in $B$ while $\Lambda$ is fixed. (Remember that $\Lambda = N + 1/2$.) For the
Bessel functions \rrf{2145} is applicable in $B$, too. Using \rrf{2430} the
asymptotics of the absolute value of the integrand in \rrf{2390} is not
greater than
    \begin{equation}
    \frac{\Lambda}{2\pi} \, e^{-2\Lambda \,
    \left[ (\ln \Lambda - \zeta) \cos \varphi -
    \varphi \sin \varphi \right] }.
    \lbq{2440}\end{equation}
If $\varphi \in B$ then $\, \cos \varphi \geq 1/\sqrt{2}\, $ and $\, \varphi
\sin \varphi \leq \pi/4\sqrt{2}\, $ so an upper estimate of \rrf{2440} is
    \begin{equation}
    \frac{\Lambda}{2\pi} \, e^{-\sqrt{2} \Lambda (\ln \Lambda - \zeta -
    \frac{\pi}{4})}
    \lbq{2450}\end{equation}
if $\ln \Lambda > \zeta$. Integrating \rrf{2450} on $B$ means simply a
factor of $\pi/2$. We see that the value of the integral tends to zero
in the limit $\Lambda \rightarrow \infty$.

In the last step let us turn to the third term of the r.h.s. of \rrf{2359}.
For great $\Lambda$-s this integral is asymptotically equal to
    \begin{equation}
    \int_C \frac{d\nu \: \nu}{i 4\sqrt{2} \sin \pi\nu}
    \left[\frac{(\frac{z_1}{2})^\nu}{\Gamma(1+\nu)}
    \frac{(\frac{z_2}{2})^{-\nu}}{\Gamma(1-\nu)}\right] +
    \left[\frac{(\frac{z_1}{2})^{-\nu}}{\Gamma(1-\nu)}
    \frac{(\frac{z_2}{2})^\nu}{\Gamma(1+\nu)}\right]
    \lbq{2460}\end{equation}
($\Lambda$ is not an integer so \rrf{2080} is applicable). Knowing that
$\Gamma(1+\nu) \Gamma(1-\nu) = \pi\nu / \sin \pi\nu$ and writing $\nu$ as
$\Lambda \exp(i \varphi)$ we get the integral
    \begin{equation}
    \frac{\Lambda}{4\pi\sqrt{2}} \int_{-\frac{\pi}{2}}^{\frac{\pi}{2}}
    d\varphi \: e^{i \varphi} \left[ e^{\frac{\Lambda}{\sqrt{2}} (x_1 - x_2)
    e^{i\varphi}} + e^{-\frac{\Lambda}{\sqrt{2}} (x_1 - x_2)
    e^{i\varphi}} \right]
    \lbq{2470}\end{equation}
(cf. \rrf{2325}). We determine this integral by expanding the exponentials in
power series and integrating the terms individually. The sum of the results is
    \begin{equation}
    \frac{\Lambda}{\pi\sqrt{2}} \sum_{k=0}^\infty (-1)^k
    \frac{\left[\frac{\Lambda}{\sqrt{2}} (x_1 -
    x_2) \right]^{2k}}{(2k + 1)!},
    \lbq{2480}\end{equation}
which is the power series of $\sin \! \left[ \frac{\Lambda}{\sqrt{2}} (x_1 -
x_2) \right] / \, \pi(x_1 - x_2)$. This function tends to $\delta(x_1 - x_2)$
if $\frac{\Lambda}{\sqrt{2}} \rightarrow \infty$, which we wanted to prove.

The method of the proof of the case $p = 2$ can be applied in a
straightforward way for any other values of $p$ as well. If $p = 1$ then a
factor of $-1$ appears at the first two terms of the r.h.s. of \rrf{2359},
which is of no significant importance in the proof. For the other possible
values of $p$ the remarkable difference is that poles come not only from $\sin
(\pi\nu)$ but also from the analytic continuation of $\exp[\pm i\theta_p(u)]$.
Nevertheless, the sum of all residues appearing in the proof gives exactly
$\,-S(x_1, x_2)$, which eliminates the first term in \rrf{2300}---just as it
happened in the case $p = 2$. Besides these extra poles the proof needs no
serious modification.

The proof given above also works for the orthogonal system \rrf{843} of the
case $s > 0$. Using \rrf{2050} the residues turn out to be zero. What is left
is just the same as we had in the case $s < 0$, $p = 1$, because of the common
asymptotic behaviour of $J_\nu(z)$ and $I_\nu(z)$ (see \rrf{2145}). This way
the case $s > 0$ can be treated with the same tools as the $s < 0$ one.

\end{document}